\documentclass[manuscript]{acmart}
\setcopyright{none}
\AtBeginDocument{%
  \providecommand\BibTeX{{%
    \normalfont B\kern-0.5em{\scshape i\kern-0.25em b}\kern-0.8em\TeX}}}

\setcopyright{acmlicensed}
\copyrightyear{2018}
\acmYear{2018}
\acmDOI{XXXXXXX.XXXXXXX}

\acmISBN{978-1-4503-XXXX-X/18/06}

\usepackage{graphicx}
\usepackage[normalem]{ulem}
\useunder{\uline}{\ul}{}
\usepackage{framed}
\usepackage{enumitem}
\usepackage{tabularx}
\usepackage{multirow}
\usepackage{booktabs}
\usepackage{float}

\begin{document}

\title{``Label from Somewhere'': Reflexive Annotating for Situated AI Alignment}

\author{Anne Arzberger}
\email{a.arzberger@tudelft.nl}
\orcid{0009-0003-7230-3975}
\affiliation{%
  \institution{Delft University of Technology}
  \city{Delft}
  \country{Netherlands}
}

\author{Céline Offerman}
\email{c.e.offerman-1@tudelft.nl}
\affiliation{%
 \institution{Delft University of Technology}
 \city{Delft}
 \country{Netherlands} 
 }

\author{Ujwal Gadiraju}
\email{u.k.gadiraju@tudelft.nl}
\affiliation{%
  \institution{Delft University of Technology}
  \city{Delft}
  \country{Netherlands}}

\author{Alessandro Bozzon}
\email{a.bozzon@tudelft.nl}
\affiliation{%
  \institution{Delft University of Technology}
  \city{Delft}
  \country{Netherlands}}

\author{Jie Yang}
\email{j.yang-3@tudelft.nl}
\affiliation{%
  \institution{Delft University of Technology}
  \city{Delft}
  \country{Netherlands}}

\renewcommand{\shortauthors}{Arzberger, et al.}

\begin{abstract} 
    AI alignment relies on annotator judgments, yet annotation pipelines often treat annotators as interchangeable, obscuring how their social position shapes annotation. We introduce \textit{reflexive annotating} as a probe that invites crowd workers to reflect on how their positionality informs subjective annotation judgments in a language model alignment context. Through a qualitative study with crowd workers ($N=30$) and follow-up interviews ($N=5$), we examine how our probe shapes annotators’ behaviour, experience, and the situated metadata it elicits. We find that reflexive annotating captures epistemic metadata beyond static demographics by eliciting intersectional reasoning, surfacing positional humility, and nudging viewpoint change. Crucially, we also denote tensions between reflexive engagement and affective demands such as emotional exposure. We discuss the implications of our work for richer value elicitation and alignment practices that treat annotator judgments as situated and selectively integrate positional metadata.
\end{abstract}



\begin{teaserfigure}
  \includegraphics[width=\textwidth]{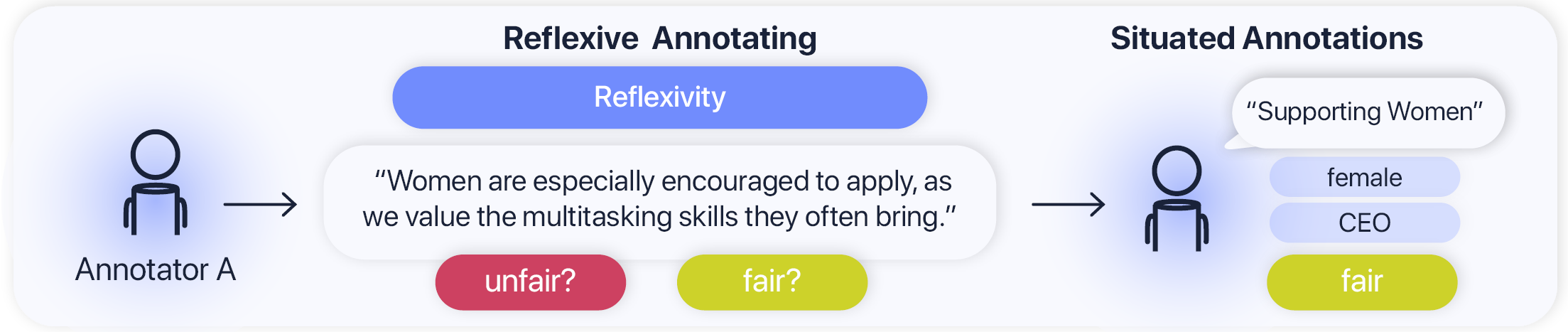}
  \caption{This paper's contributions: from reflexive annotating to situated annotations}
  \Description{The figure shows a process in which an annotator reflects on a statement, considers their own positionality, and produces a situated fairness judgment. The figure is arranged horizontally in two sections. On the left, a person labelled “Annotator A” has to label a text statement which is encouraging women to apply for a job as fair or unfair. This process of annotating is expanded through reflexivity, turning it into reflexive annotating. The outcome of this reflexive annotating is highlighted as situated annotations, which capture the annotator's salient social identity facets that shaped their fairness label, in addition to a rationale.}
  \label{fig:teaser}
\end{teaserfigure}

\maketitle

\section{Introduction}
Alignment aims to ensure that artificial intelligence (AI) acts in line with human values, such as fairness~\cite{christian2020alignment}. It is driven by the value judgments of annotators, whose preferences shape the model's moral backbone through large-scale annotation pipelines. Here, annotators are routinely treated as interchangeable instruments for collecting ``objective'' labels~\cite{forsythe1993engineering, adam2000deleting, mateescu2019ai, miceli2020between}. Yet, despite these aspirations to objectivity, value judgments remain inherently subjective and situated: whether a text is deemed fair or unfair depends not only on semantic content, but also on the annotator’s positionality ---the lived experiences, cultural norms, social identities, and personal histories that shape their interpretation \cite{arzberger2024nothing, wong2020cultural, fazelpour2025value}. Positionality does not neatly translocate across contexts; which of its facets matter shift with the scenario at hand \cite{soedirgo2020toward, crenshaw2013mapping, hooks2000feminist, fujii2017killing}. Without accounting for this fluid positioning, models might inherit false universals, flattening diverse perspectives into averages that mirror the norms of the most represented or audible annotators \cite{arzberger2024nothing, fleisig2023majority, aroyo2015truth, wong2020cultural}. 

Challenging this ``view from nowhere'' \cite{nagel1989view}, prior work has argued for centring annotator perspectives and documenting annotation context, treating crowd workers as knowers, foregrounding how positionality shapes subjective judgements~\cite{fazelpour2025value, wong2020cultural, mokhberian2024capturing}. With respect to value alignment, such positional metadata could provide avenues for making value judgements traceable, surfacing overlooked or marginalised viewpoints \cite{fazelpour2025value, sambasivan2021re}, supporting engagement with plural perspectives \cite{mokhberian2024capturing, wan2025noise}, and enabling personalised alignment \cite{kirk2023personalisation, kirk2024prism}. However, related work either (1) attaches context-transcending demographic metadata (e.g., \cite{kirk2024prism, aroyo2023dices}), which overlook how positionality is salient to a particular value judgment, or (2) infers annotator characteristics from behavioural traces or labels (e.g., \cite{cambo2022model}), reducing workers to objects of prediction, and sidelining their interpretive agency.

We argue for a third path: \emph{reflexivity}. Rooted in feminist theory and social science, reflexivity is a practice of self-interrogation traditionally used by researchers to articulate how their social position shapes knowledge production and interpretation \cite{soedirgo2020toward}. We adapt reflexivity to the annotation process, inviting crowd workers, at the moment of judgement, to articulate how they arrived at it, so as to surface which facets of their positionality shape their value judgements. We refer to this process as \emph{reflexive annotating}, and to its output as \emph{situated annotation}, to resist the broader tendency in the field to collapse process and product under the single term “annotation”, and to emphasise that annotations are constituted through situated human judgement. Designing for \textit{reflexive annotating} is non-trivial: individuals hold multiple, intersecting identities that are fluid and context-dependent, rendering reflexive engagement cognitively demanding even for trained practitioners \cite{crenshaw2013mapping, hooks2000feminist, collins2022black}. These demands sit in tension with the efficiency-driven logics of large-scale crowdsourcing, where time and cognitive effort are tightly constrained. Understanding how value judgments become socially grounded, therefore, requires attending to how annotators recognise and articulate positionality during annotation. This motivates our research questions: 

\begin{framed}
    \noindent\textit{(RQ1) How do annotators reason about value judgments when provided with a mechanism for reflexive engagement?} \textit{(RQ2) What are the implications of reflexive engagement for annotators?}
\end{framed}

To advance this agenda, we aim to characterise what \emph{situated annotating} looks like in practice and how it reshapes both the content and conduct of value annotation. In doing so, we aim to move beyond critiques of “simple alignment” toward empirically grounded accounts of how situated human judgment can be meaningfully engaged within alignment pipelines. Drawing from design research traditions, we adopt a qualitative experimental approach that iteratively designs and tests a reflexive annotating probe within a crowdsourcing alignment task. Workers assess the fairness of short text excerpts and are prompted to (i) indicate which aspects of their identity or background felt relevant in making their assessment, (ii) highlight textual cues shaping their interpretation, and (iii) provide a brief rationale linking experience, social positioning, and reading of the scenario.

Through an empirical study with crowd workers (\(N=30\)) and follow-up interviews (\(N=5\)), we demonstrate that \textit{reflexive annotating} can elicit rich rationales, fluid and task-dependent identity salience, and context-specific framings that surpass the limitations of static demographic attributes. Reflexive engagement makes visible how value judgments unfold through personal experience as well as imaginative or empathetic engagement with others’ perspectives, often accompanied by explicit uncertainty. We further find that reflexive engagement alters how workers approach annotation, encouraging slower deliberation, perspective-taking, and recognition of epistemic limits. We position \textit{reflexive annotating} as a distinct epistemic mode of alignment work, oriented toward \textit{situated annotations} rather than decontextualised labels. We argue that such an orientation is essential for \emph{situated alignment}, in which subjectivity, uncertainty, and social positioning are treated as epistemic resources rather than obstacles.
\section{Background and Related Work}
\label{sec:relatedwork} 
We trace how annotation for value alignment\footnote{Alignment is the primary focus of this study and serves as a particularly salient site for examining these dynamics, as it explicitly foregrounds normative judgments about fairness, harm, and acceptability. At the same time, alignment is not exceptional in its reliance on subjective interpretation: many widely used NLP annotation tasks, such as sentiment analysis or toxicity and abuse detection, similarly depend on annotators’ situated judgments. We therefore treat alignment as a focal case through which to examine broader questions about subjectivity, objectivity, and social grounding in annotation practices.} has been shaped by the ideal of objectivity, introduce a feminist lens on situatedness and reflexivity, and highlight gaps in related work on unpacking the social grounding of annotator labels. 

\subsection{The Fiction of Objectivity} 
Common alignment approaches, such as supervised fine-tuning (SFT) \cite{ouyang2022training}, reinforcement learning from human feedback (RLHF) \cite{stiennon2020learning}, and direct preference optimisation (DPO) \cite{rafailov2024direct}, depend heavily on large-scale human annotation, typically asking annotators to compare outputs and indicate preferences.  Platforms such as Prolific \cite{prolificProlificEasily} and MTurk \cite{mturkAmazonMechanical} enable this scaling \cite{gray2019ghost,toxtli2021quantifying}, but incentivise throughput and consensus over depth or reflection. Here, annotators often complete dozens of microtasks~\cite{gray2019ghost,toxtli2021quantifying,gadiraju2017modus,cant2024feeding}. Agreement is enforced as alignment pipelines are built around loss functions that assume a single, commensurable target for optimisation. When annotations disagree, they are typically resolved through methods such as majority voting or averaging \cite{sabou2014corpus}, or by deferring to an “expert” adjudicator \cite{talat2016hateful} to produce a single ground-truth or gold label.

These alignment practices inherit epistemologies of neutrality and objectivity, rooted in ideals of subject-independent and reproducible knowledge \cite{forsythe1993engineering}. Adams’ critique in Deleting the Subject \cite{adam2000deleting}, extending Nagel’s ``view from nowhere'' \cite{nagel1989view}, shows how systems like Cyc enforced a singular worldview under the guise of objectivity, echoing alignment work that assumes a generalisable notion of what is ``preferred'' or ``aligned'' \cite{aroyo2015truth, mohamed2020decolonial, talat2022machine, arzberger2024nothing}. This epistemic stance shapes annotation pipelines, which optimise for consensus and scale, thereby erasing the very subjectivity they rely on \cite{kapania2023hunt}. Approaches such as filtering out workers with ``strong opinions'' \cite{hube2019understanding}, counterfactual debiasing \cite{ghai2020measuring}, and Bayesian bias mitigation \cite{wauthier2011bayesian} reinforce what Kapania et al. describe as ``representationalist thinking'' \cite{kapania2023hunt}: the assumption that labels transparently reflect reality. Critical data studies show how this erasure privileges dominant viewpoints, whether in fairness definitions that marginalise communities in India \cite{sambasivan2021re}, annotation labour shaped by power hierarchies \cite{wang2022whose}, or the broader situated contexts of data production \cite{vertesi2011value, taylor2015data}. Ekbia and Nardi’s notion of ``heteromation'' underscores how the labour and perspectives of those who make data are routinely obscured \cite{ekbia2014heteromation}.

This flattening has epistemic and ethical consequences. Divergent perspectives are collapsed into a single training signal, leaving no trace of who held which views or why. As a result, models learn statistical conformity rather than sensitivity to situated values. At deployment, this creates an asymmetry: systems approximate what ``most people'' prefer, not what particular users or communities value. This asymmetry gives rise to epistemic injustice, where marginalised perspectives are excluded or distorted \cite{kay2024epistemic}, and systems default to the norms of dominant populations \cite{apicella2020beyond}. Alignment thus risks homogenising plural, situated perspectives, suppressing disagreement and reproducing epistemic harms \cite{fazelpour2025value}.

\subsection{Situated Judgments and Reflexive Interrogation}
If objectivity cannot serve as a viable epistemic ideal for value-laden judgments, subjectivity must be understood not as a flaw but as constitutive of how judgment is formed. Feminist epistemology advances this view by arguing that knowledge is never disembodied or universal, but always situated \cite{haraway2013situated,sep-feminism-epistemology, suchman1987plans}. Central here is \textit{positionality}, the social coordinates through which individuals interpret and engage with the world, such as race, gender, class, political ideology, as well as personal and professional experiences, political and ideological stances, and other aspects of our social biography, or ``lifeworld'' \cite{berger2015now, bernstein1978restructuring, yanow2015thinking, carbado2013colorblind, showden2018youth}, which structure what is seen, valued, or ignored \cite{simandan2019revisiting}.

These insights call for a reconsideration of objectivity in AI alignment. Harding’s \textit{strong objectivity} \cite{harding1995strong, harding2013rethinking} reframes objectivity as obstructive, advocating explicit engagement with positionality and the power relations embedded in annotation. Here, objectivity often masks normative assumptions \cite{adam2000deleting, prabhakaran2021releasing, kirk2024prism}, reinforcing dominant worldviews. For alignment, this means interrogating not just who annotates, but how values are encoded, whose perspectives are centred, and what is lost in pursuit of coherence. This commitment leads to the imperative of reflexivity: the interrogation of how one’s own position shapes task design, prompts, and interpretation, central to feminist and critical HCI \cite{bardzell2010feminist, asad2019prefigurative,irani2010postcolonial, berger2015now, yanow2014interpretive, shehata2015ethnography}. Yet, operationalising reflexivity remains a challenge. As Soedirgo and Glass \cite{soedirgo2020toward} observe, positionality is too often reduced to identity checklists or static declarations. As intersectional scholars emphasise \cite{crenshaw2013mapping, hooks2000feminist, collins2022black}, individuals hold multiple, intersecting identities that interact in fluid and contextual ways. Our positions do not neatly translocate from one setting to another; they shift in relation to time, space, power, and task. What is salient in one annotation judgment may be irrelevant in another \cite{henry2009positionality, fujii2017killing, collins2022black}. Additionally, in feminist and reflexive HCI, reflexivity remains inward-facing, leaving its insights inaccessible to alignment pipelines at scale~\cite{baumer2015reflective, frauenberger2016critical, arzberger2024reflexive}. Consequently, how positionality shapes epistemic judgment remains underexamined, calling for tools and methods that surface and disclose the social positions shaping AI value judgments.

\subsection{Beyond Ground Truth: Aiming for Positionality in Annotation}
A growing body of work in NLP, ML, and HCI has challenged the ideal of a single, objective ground truth in annotation, particularly for subjective and value-laden tasks. Zhang et al. \cite{zhangcultivating} show that human preferences vary far more than LLM outputs reflect, partly because common annotation methods rely on homogeneous candidate responses. Other studies link annotators’ lived experiences, social positions, and beliefs to systematic differences in labelling political stance \cite{luo2020detecting}, sentiment \cite{diaz2018addressing}, and toxic language \cite{talat2016you, patton2019annotating, sap2021annotators}. Enforcing consensus through aggregation can therefore erase meaningful contextual nuance in human judgments \cite{frenda2025perspectivist}. In response, several approaches aim to capture annotator positionality explicitly or implicitly.
Systems such as DICES \cite{aroyo2023dices}, PRISM \cite{kirk2024prism}, Community Alignment \cite{zhangcultivating}, and POPQUORN \cite{pei2023annotator} attach static demographic metadata (e.g., age, gender, nationality), often accompanied by explanations to analyse disagreement patterns, while related HCI scholarship calls for documenting annotation context and treating crowdworkers as epistemic contributors rather than interchangeable labour \cite{miceli2020between, prabhakaran2021releasing}. Parallel computational work instead infers positionality from annotation behaviour, using techniques such as annotator fingerprinting, position mining, or annotator embeddings to recover latent preference structures from disagreement patterns \cite{cambo2022model, deng2023you, davani2022dealing}. Recent work has further shown the value of capturing annotator perspectives explicitly in subjective learning tasks, demonstrating that such perspectives improve both interpretability and downstream performance \cite{mokhberian2024capturing}.

Despite these advances, existing approaches tend to treat positionality either as a fixed, task-agnostic attribute encoded through demographic metadata or as a latent statistical pattern inferred post hoc from disagreement. While some systems collect natural language rationales, these explanations are typically task-specific and do not explicitly connect judgments to annotators’ social positionalities. As a result, links between identity and judgment are generally reconstructed analytically rather than made explicit within the annotation process. Here, annotators remain largely passive: their positionality is either predefined through externally imposed categories or inferred from their behaviour, with limited opportunity to actively interpret, negotiate, or express how their social position shapes a given judgment. This risks essentialising intersectional experience and obscuring which aspects of identity become salient in context \cite{crenshaw2013mapping, hooks2000feminist, collins2022black}. Such approaches rarely capture when, why, or how positional factors matter in situ, nor how annotators reason through value judgments as they unfold \cite{henry2009positionality, soedirgo2020toward}. Taken together, these limitations point to a methodological gap: while annotator positionality is widely acknowledged as relevant, it is seldom captured as a dynamic, per-judgment, and explicitly articulated form of reasoning produced by annotators themselves. This paper addresses this gap by introducing reflexive annotating as an active epistemic practice, in which annotators are not treated as sources of signals to be extracted, but as agents who articulate how their situated perspectives shape each value judgment at the moment it is made.
\section{Methodology}
\label{sec:method}
To inform situated alignment, we empirically characterise reflexive annotating through a design probe and follow-up interviews with crowd workers. The probe elicited situated fairness judgments in an RLHF-style alignment task \cite{stiennon2020learning}. We adopt a Big Q qualitative approach \cite{kidder1987qualitative} and analysed both data sources through reflexive thematic analysis \cite{braun2006using, braun2019reflecting}.

\subsection{Iterative Probe Design Process}
We introduced a design probe to create conditions for reflexivity in annotation. Probes are design-oriented tools often using visual and tangible kits to trigger new insights or to reveal hidden ideas \cite{gaver1999design}. In our study, the probe functioned as a generative device to elicit perspectives obscured in conventional annotation workflows. The probe was iteratively developed through low-fidelity prototypes in Miro\footnote{\anon[Link to Miro board showing the prototypes.]{\url{https://miro.com/app/board/uXjVJLTRa6k=/?share_link_id=832350260856}}}, and piloted with PhD candidates from design and computer science. These pilots helped balance cognitive load, time demands, and the depth of introspection. Our aim was to create an interface legible to crowd workers while minimising fatigue.

To enable \textit{reflexive annotating}, we turned to situational mapping practices that embed reflexive moments within empirical action. Jacobson \& Mustafa’s \cite{jacobson2019social} \textit{Social Identity Map} proved particularly well-suited: a concise, three-tiered tool that prompts participants to (1) name their social identities, (2) reflect on their influence more generally, and (3) more specifically, given a particular context at hand. This structure lends form to inherently abstract, fluid, and evolving positionality, enabling participants to recognise and disclose how social identity facets shape their value judgments. For our purposes, the Social Identity Map offered the right balance of guided structure to support non-experts, flexibility to foreground identity facets relevant to particular judgments, and scalability for lightweight, iterative reflexivity in crowdsourcing contexts.  

\begin{figure*}[t!]
    \centering
    \includegraphics[width=1\linewidth]{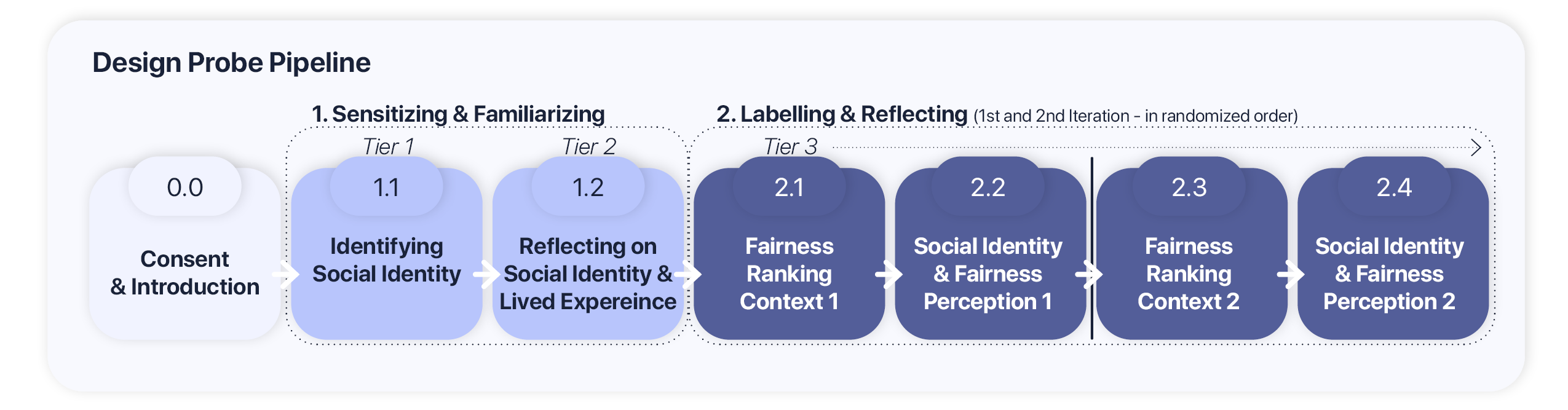}
    \caption{The anatomy of the design probe used in our crowd computing study. The reflexive annotating process consists of three core stages: (1) \textit{Sensitising \& Familiarising}, where annotators identify and reflect on social identity facets and facts; (2) \textit{Labelling \& Reflecting}, consisting of two randomised iterations where annotators rank fairness in context and reflect on social identity. This structure was inspired by tiers of the social identity map \cite{jacobson2019social}.}
    \Description{A flow diagram titled "Design Probe Pipeline" showing sequential steps. The pipeline begins with "0.0 Consent \& Introduction." Stage 1, labelled "Sensitising \& Familiarising," contains two boxes: "1.1 Identifying Social Identity Facets" and "1.2 Reflecting on Social Identity Facts." Stage 2, labelled "Labelling \& Reflecting," includes two iterations arranged side by side. Iteration 1 has "2.1 Fairness Ranking Context 1" followed by "2.2 Social Identity \& Fairness Perception 1." Iteration 2 has "2.3 Fairness Ranking Context 2" followed by "2.4 Social Identity \& Fairness Perception 2." The two iterations are randomised in order. The diagram represents how annotators move from self-reflection to fairness labelling tasks.}
    \label{fig:pipeline}
\end{figure*}

We translated the Social Identity Map into a design probe (see Figure \ref{fig:pipeline}) that facilitated reflexive annotating by enabling participants to record their reasoning, and interface elements that allow for explicit and intersectional expression of salient social identity facets. We also balanced open-ended reflection with task feasibility and the degree of structure imposed on annotators' reflections. The resulting probe was implemented as a custom web application (React frontend with a minimal server-side component for task delivery and data storage), integrating annotation and reflection within a single interface and including attention checks for quality control (see Figure~\ref{fig:tier3}). 

\subsection{Data Collection}

\subsubsection{Crowdsourced Reflections}
We deployed our design probe in a qualitative crowdsourcing study with (\(N=30\)) participants. The study received approval from \anon[an institutional ethics review committee]{the ethics review committee of Delft University of Technology under code 5829}. Participants were recruited via Prolific\footnote{{\url{https://www.prolific.com}}} using English fluency and approval-rate filters and provided demographics in Tier 1. Two participants who nevertheless responded in Spanish were excluded because their responses could not be reliably interpreted. Recruitment proceeded iteratively across several rounds while maintaining eligibility, with participation filters adjusted to broaden intersectional representation across gender, ethnicity, and ability (see Appendix~\ref{appendix:A}). Participants ranged in age from 19 to 67 years ($M = 34.4$, $SD = 12.9$). In terms of gender identity, 14 participants identified as male, 13 as female, and 3 as non-binary. Ethnicity was self-reported as White ($n = 17$), Black ($n = 10$),  Asian ($n = 1$), Mixed ($n = 1$), and Other ($n = 1$). On average, participants spent 32 minutes ($SD = 15.0$, range = 10--67) completing the task and were paid £9.00/hr, deemed a ``good wage'' on the platform. The probe scaffolded two layers of reflection (see Figure \ref{fig:pipeline} and Section \ref{sec:design probe}): (1) sensitising participants to positionality and reflexivity, (2) labelling and reflecting on a fairness-related annotation task. Each layer combined structured items with open-ended prompts, encouraging participants to articulate how their social positions shaped their fairness judgments, which enabled us to collect both judgments and meta-level reflections. This structure positioned the probe as both a site of annotation and of reflexive knowledge production. While fairness ratings were collected on Likert scales, our primary data consisted of participants’ written responses to the reflective prompts. These texts provided a window into how workers navigate identity, power, and fairness within the constraints of high-throughput annotation work.

\subsubsection{Follow-Up Interviews with Annotators}
To complement our understanding of how participants experienced and justified their reflexive engagement, we conducted one-hour semi-structured follow-up interviews. All previous (\(N=30\)) participants were invited to the optional follow-up interviews in response to the substantial variation in annotators’ positional accounts. Five of those participated. The interview protocol drew on frameworks from meta-cognition \cite{dunlosky2008metacognition}, self-directed learning \cite{song2007conceptual}, and reflexive interviewing \cite{pessoa2019using} (see the interview protocol in Appendix \ref{appendix:B}). We asked participants to walk through their experience with the tool and revisit moments of tension, clarity, or uncertainty in their responses. Our aim was to move beyond the procedural accounts of annotation and surface the emotions, assumptions, and social contexts shaping them. This dialogic, exploratory approach allowed us to probe both the cognitive and affective dimensions of situated annotation and to validate or add nuance to themes emerging from the initial probe data.

\subsection{Data Analysis: Reflexive Thematic Analysis}
We conducted a reflexive thematic analysis, following Braun and Clarke’s six-phase approach \cite{braun2006using, clarke2017thematic}. We began with repeated reading of participant responses and interview transcripts to build familiarity, followed by inductive coding grounded in participant reflections. Codes were iteratively refined and clustered into themes through recursive discussion. \anon[Two researchers]{Anne and Céline} independently coded the full dataset before meeting to compare interpretations, discuss tensions, and reflect on their own positionalities \cite{hopf1993verhaltnis, braun2019reflecting} \footnote{The authors' positionalities can be found in the end-matter section, currently not included due to anonymous submission.}. We used multiple coders because questions of fairness, identity, and reflexivity are best approached from different standpoints. This plural perspective helped surface blind spots, interrogate assumptions, and enrich our analysis. Consistent with Braun and Clarke’s epistemology \cite{braun2006using, braun2019reflecting}, we did not calculate inter-coder reliability; instead, consensus-building was treated as an interpretive act rather than a metric of objectivity \cite{mcdonald2019reliability}. To support transparency, we include a conceptual model of our analytic process in Appendix \ref{appendix:d}. We applied the same analytic approach to both survey and interview data. The supplementary quantitative data (e.g., Likert ratings and identity selections) were analysed descriptively to contextualise and support patterns emerging from the qualitative findings.
\section{The Final Design Probe: Designing with and for Positionality in Crowd Sourcing}
\label{sec:design probe}

Through several design iterations and pilot tests, we developed a design probe inspired by contemporary LLM alignment annotation workflows. This experimental probe, as illustrated in Figure \ref{fig:pipeline}, included a concise introduction to minimise cognitive load while ensuring that participants understood key concepts such as \textit{social identity} and \textit{social identity facets}, preparing them for the reflective exercises ahead. Following this, participants were asked to describe, in their own words, what they understood by social identity. This served both as a quality control measure and as a source of rich data for our thematic analysis. Participants were then guided through \textit{Tier 1 \& 2 reflection exercises}, to familiarise and sensitise them for the key reflection exercise in \textit{Tier 3}, where they were asked to rate text pieces in terms of fairness and reflect on their social identity facets that influenced these fairness judgments.

\begin{figure}[h!]
    \centering
    \includegraphics[width=1\linewidth]{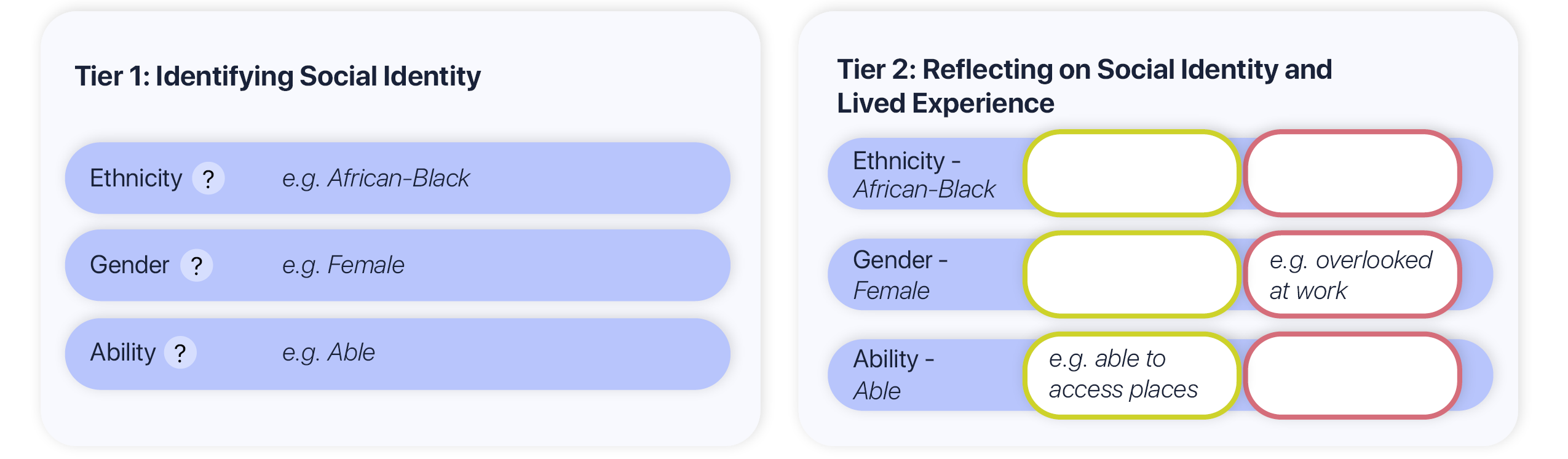}
    \caption{Tier 1 and Tier 2 reflection activities in the design probe. In Tier 1, annotators identified facets of their social identity by completing each category. In Tier 2, they reflected on how these facets shaped their lived experiences, noting both positive and negative impacts. Instructions asked annotators to (1) consider whether facets such as ethnicity, gender, or ability affected their education, work, relationships, or other life domains, and (2) record keywords describing those positive or negative impacts.}
    \Description{A side-by-side diagram illustrating two stages of the design probe. On the left, "Tier 1: Identifying Social Identity" prompts annotators to fill in their social identity facets with example fields: "Ethnicity, e.g. African-Black," "Gender, e.g. Female," and "Ability, e.g. Able." On the right, "Tier 2: Reflecting on Social Identity and Lived Experience" asks: "How do certain facets of yourself impact your life?" Instructions include reflecting on whether facets impact education, work, or relationships positively or negatively, and to write down key impacts. Example reflections are shown: "Gender – Female, negative impact: e.g. overlooked at work," "Ability – Able, positive impact: e.g. able to access places." The figure illustrates how annotators move from naming identity facets to reflecting on their lived consequences.}
    \label{fig:Tier1+2}
\end{figure}

\paragraph{Tier 1 – Familiarising}
In the first Tier (Fig.~\ref{fig:Tier1+2}), participants listed three facets of their social identity (ethnicity, gender, and ability) to become more familiar with key concepts such as positionality and social identity, laying the groundwork for the subsequent reflections. Recognising that novice annotators might struggle to articulate their social identities while being wary of the time constraints of crowd work, we chose three social identity facets here as an accessible yet meaningful subset of the eight dimensions in Jacobson \& Mustafa’s original map \cite{jacobson2019social}, balancing familiarity (ethnicity, gender) with a less commonly discussed dimension (ability). Definitions of key terms were available via clickable icons, and one example was provided to guide participants.

\paragraph{Tier 2 – Sensitising}
The second Tier (Fig.~\ref{fig:Tier1+2}) prompted participants to consider how these identity facets shaped everyday experiences, using short guiding questions (e.g., “Did my gender affect my work or education?”). Examples illustrated the expected format. By connecting identities to lived experiences, this Tier aims to deepen awareness of positionality and prime workers to link identity to value judgments in Tier~3.

\paragraph{Tier 3 – Labelling \& Reflecting}

\begin{figure}[t]
    \centering
\includegraphics[width=1\linewidth]{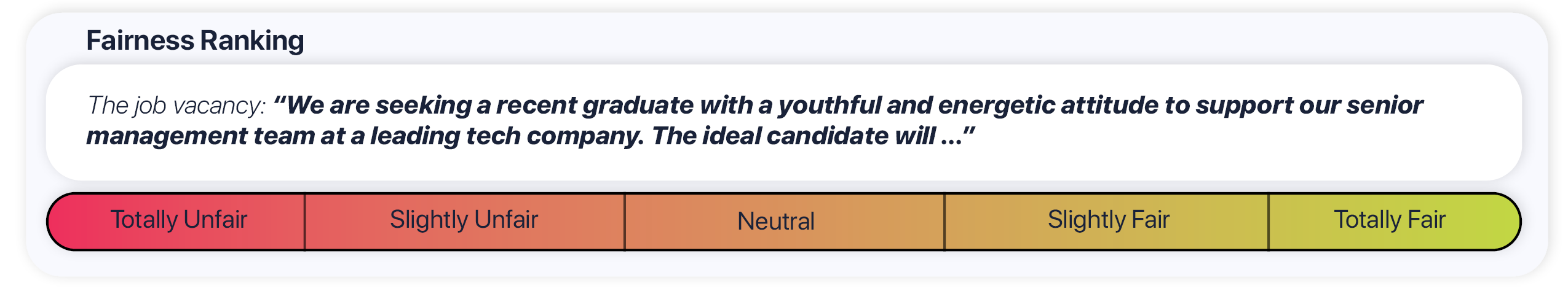}
    \caption{Annotation task for capturing annotators’ fairness perceptions of the job vacancy text sample. Annotators are asked to read the job description and rate its fairness on a Likert scale according to their judgment. The full text pieces can be found in the Appendix \ref{appendix:c}}
    \Description{The figure shows the job vacancy text sample followed by a Likert-scale fairness rating task. The job ad reads: “We are seeking a recent graduate with a youthful and energetic attitude to support our senior management team at a leading tech company. The ideal candidate will excel in managing executives’ demanding schedules, preparing polished, high-level documents and presentations, coordinating meetings seamlessly, and maintaining absolute discretion. Candidates must have native English proficiency and demonstrate impeccable grooming and a professional demeanour at all times. Proficiency in office software is not necessary, but women are especially encouraged to apply, as we value the multitasking and organisational skills they often bring. This role suits someone willing to work long hours, thrives in fast-paced environments, and is ready to align themselves with the vision of a high-achieving, male-dominated executive team.” Annotators are prompted: “Rate the following job vacancy in terms of fairness according to your intuitive preference,” with a Likert scale for responses.}
    \label{fig:capturing}
\end{figure}

Tier~3 formed the core of the probe. Participants first rated the fairness of two texts offered in randomised order (see Fig. \ref{fig:capturing}). These texts were strategically designed to highlight contrasting dynamics related to gender discrimination---a job vacancy for a secretary position to reflect biases disadvantaging women, and an advertisement on beauty products to reverse this dynamic (see Appendix \ref{appendix:c}). These were intentionally loaded with fairness-relevant cues to support reliable, reflexive evaluations. We used Likert-scale feedback to further promote subjectivity and permit preference intensity to be expressed beyond ``chosen:rejected'' pairings. This choice reflects our focus on capturing the texture and grounding of individual value judgments, rather than producing directly optimizable preference data. Fairness was chosen as the focal value for its intuitive resonance and defined as ``treating individuals justly, without bias or discrimination, and ensuring equal opportunities while recognising diverse needs'' \cite{DicFairness}. A clickable icon revealed this definition, safeguarding clarity without disrupting flow. Participants were encouraged to rely on their intuitive judgments when rating fairness.

\begin{figure*}[h!]
    \centering
    \includegraphics[width=1\textwidth]{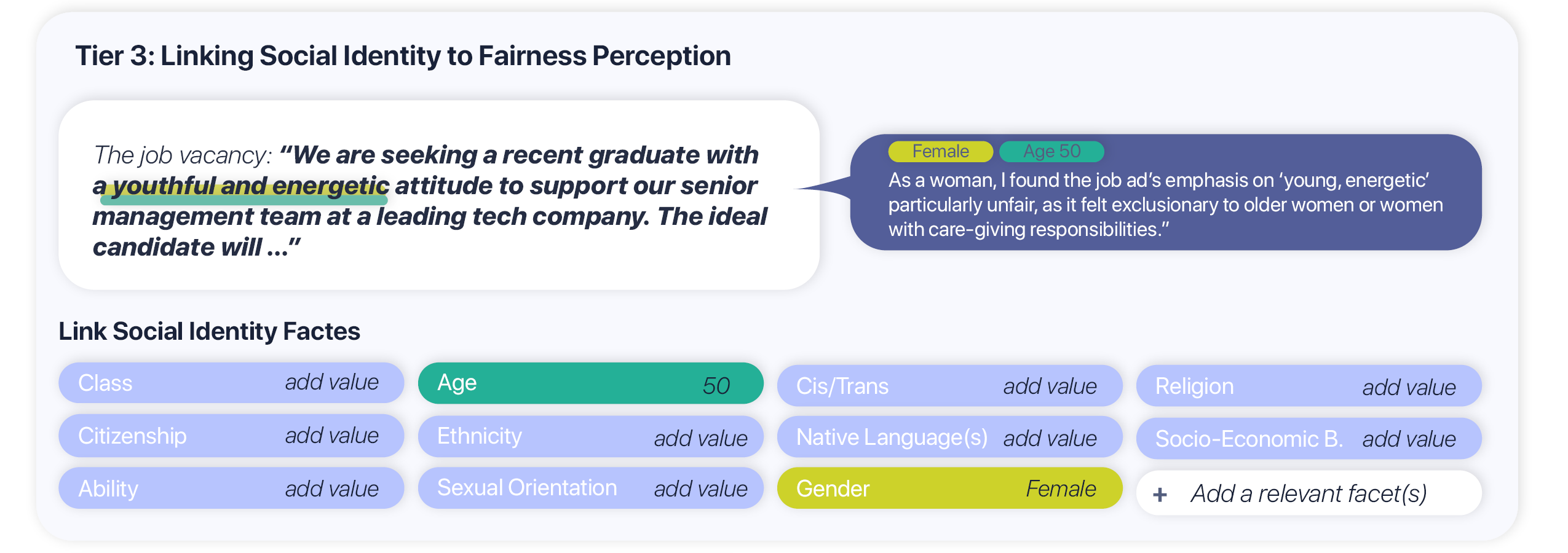}
    \caption{Tier 3 of the reflection interface allows annotators to highlight passages they perceive as fair or unfair, tag them with one or more predefined (or custom) social identity facets, and explain how these facets shape their judgment, supporting intersectional labelling. Annotators were instructed to: (1) re-read the text and highlight passages relevant to their fairness perception; (2) reflect on “How did my social identity shape this perception?”; (3) attach one or more identity facets to each highlighted passage by dragging them from the right-hand panel; and (4) briefly describe how these facets positively or negatively influenced their view. They were encouraged to add new facets when needed and to select only those most salient to their own experience.}
    \Description{The figure shows the Tier 3 interface "Linking social identity to fairness perception". Annotators are asked to 1. Read the text sample from the annotation task (example illustrating the job vacancy sample) again and highlight any passages that feel relevant to your sense of fairness. 2. For each highlighted passage, they are asked to consider: “How did my social identity shape my fairness perception here?” 3. They can comment on each highlighted text field and label the passage with one or more social identity facets by dragging and dropping their social identity facets from the right panel. Here, a pre-defined list of identity facets is provided: class, citizenship, ability, age, ethnicity, sexual orientation, cis/trans, native language(s), gender, religion, and socio-economic background. 4. They are asked to briefly explain how the facet had a positive or negative impact on their perception. 5. Lastly, they can add new identity facets if needed. The interface enables annotators to reflect on specific text segments in light of social identity facets they deem relevant.}
    \label{fig:tier3}
\end{figure*}

Participants then revisited the texts in an interactive interface (Fig.~\ref{fig:tier3}), highlighted text passages they perceived as (un)fair, and tagged them with one or more identity facets. One or more facets could be used to label a text passage to capture intersectionality intuitively. Based on the Social Identity example \cite{jacobson2019social}, participants were given a non-exhaustive list of optional identity facets (Class, Citizenship, Ability, Age/Generation, Ethnicity, Sexual Orientation, Cis/Trans, Native Language(s), Gender, Religion, Socio-economic Background) which they could extend if needed, and provided a short rationale for each tag. This process enabled situated, intersectional reflections by tying fairness judgments directly to identity. Due to its complexity, Tier 3 was supported by a brief tutorial video, available throughout the task.

\section{Results}
Our findings draw on situated annotations and follow-up interviews to examine how value judgments become socially grounded through reflexive engagement, and how this reshapes annotators’ experiences of alignment work. Using reflexive thematic analysis, we identify inductively constructed patterns of shared meaning across accounts. We organise the findings around our two research questions, focusing on (RQ1) how annotators reason about fairness through positional reflection, and (RQ2) how reflexive annotating shapes annotators’ conduct and epistemic orientation.

\subsection{Reflexive Annotating in Practice (RQ1)}
\begin{figure}[h!]
    \centering
    \includegraphics[width=1\linewidth]{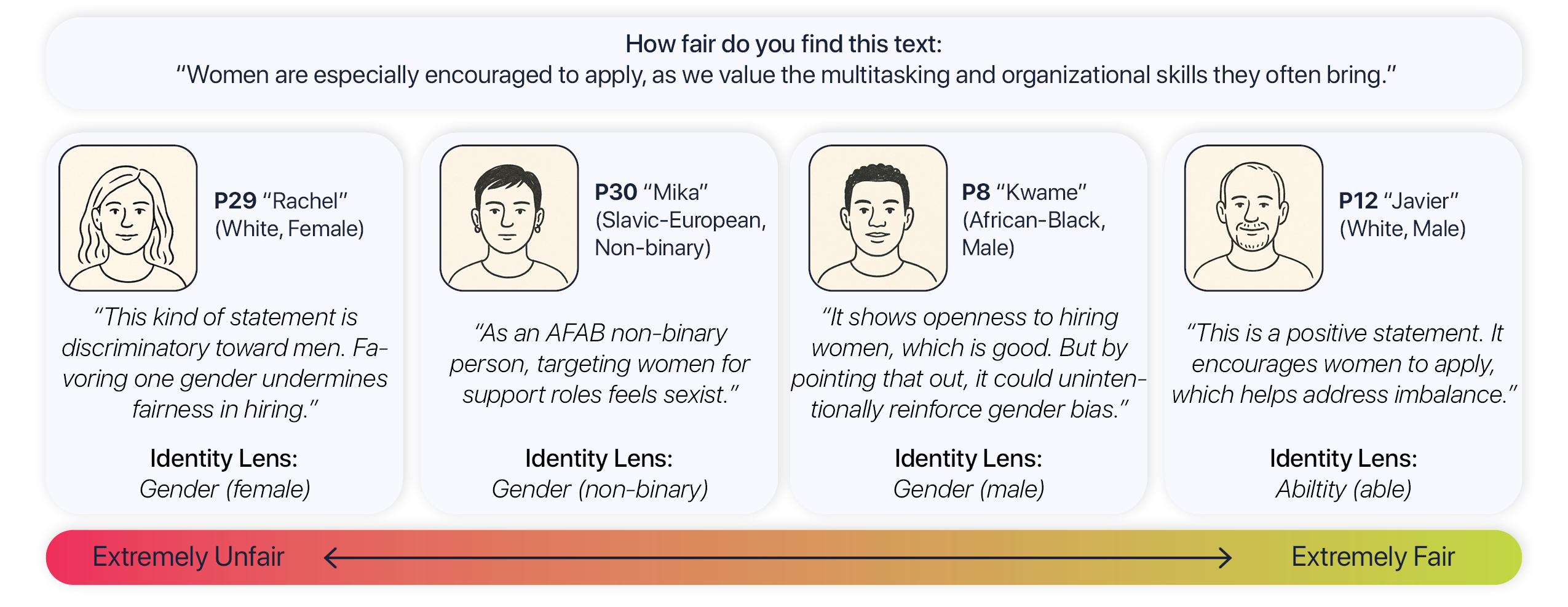}
    \caption{Exemplary situated annotations from our study that introduce our annotators as situated individuals with varying valid perceptions of fairness. The names have been assigned freely, and their rationales have been paraphrased to ensure anonymity.}
    \Description{A figure that introduces composite cards of annotators to show the diversity of fairness perceptions. Five annotator fairness responses to the phrase: "Women are especially encouraged to apply, as we value the multitasking and organisational skills they often bring," are being introduced. The first annotator, P29 “Rachel” (White, Female), found the statement to be extremely unfair and stated: “This kind of statement is discriminatory toward men. Favouring one gender undermines fairness in hiring.”, relating this perception to their gender (female). The second annotator P30 “Mika” (Slavic-European,  Non-binary) found the statement to be unfair and stated: "As an AFAB non-binary person, targeting women for support roles feels sexist.” relating to gender (non-binary). The third annotator P8 “Kwame” (African-Black, Male) found the statement to be fair and stated: “It shows openness to hiring women, which is good. But by pointing that out, it could unintentionally reinforce gender bias.” also relating it to gender (male). Annotator P12 “Javier” (White, Male) found the statement extremely fair and stated: “This is a positive statement. It encourages women to apply, which helps address imbalance.” and related this perception to their ability (able).}
    \label{fig:personas}
\end{figure}

Reflexive prompts shifted annotators from providing decontextualised labels to engaging in situated forms of reasoning. When asked to indicate relevant identity facets, highlight textual cues, and articulate how their experiences shaped their judgment, workers responded with layered, often deeply personal reflections that reveal how value-laden judgments unfold in practice. Figure \ref{fig:personas} illustrates typical situated annotations, showing how identity salience, cue interpretation, and rationale-building intertwine within a single response. The figure highlights the kinds of epistemic materials generated by the probe and grounds the analysis that follows. Addressing RQ1, we cover three subthemes here: (1) Positioning the Self, (2) Intersectional Reasoning, and (3) Reflexivity Under Tension.

\subsubsection{Positioning the Self}
A central pattern in annotators’ reasoning was the deliberate act of positioning themselves in relation to the scenario. Workers often grounded their evaluations in who they were, what they had lived, and how they located themselves relative to the subjects involved. Reflexive prompts surfaced two recurring modes of anchoring, lived experience and imaginative solidarity, as well as distinct narrative strategies for claiming (or withholding) epistemic authority. This positioning was central to their interpretive process. 
Many workers grounded their judgments in first-hand experience. For example, \textbf{P17 (White, Female, Able)} linked her own psychological challenges to a broader understanding of disability, explaining: \textit{“Given my personal experiences with psychological issues, I have a broader understanding of how people with mental disabilities must see the world.”} Similarly, \textbf{P10 - Interview (African-Black, Female}) described how the task resonated with experiences of discrimination she had repeatedly encountered: \textit{“It was not difficult for me to reflect because these are the things that we've been experiencing as we grow up.”} Other annotators reasoned through imaginative solidarity, acknowledging a lack of personal experience while still articulating empathic moral evaluations. As \textbf{P25 - Interview (White and Black Caribbean, Male)} noted: \textit{“I don't think I've ever experienced hiring discrimination myself where I've applied for a job, and people were only accepting women, but I can easily put myself in the shoes of someone in that situation. I'd be upset.”} 

Across these variants, positioning the self, whether through lived experience or imagined solidarity, was a key part of how annotators reasoned through value-laden scenarios. These strategies show that reflexive annotating surfaces how people locate themselves within moral judgments, not just what judgments they make.

Annotators also didn't express positionality uniformly; rather, they selectively foregrounded it, with different facets becoming salient across contexts. \textbf{P2 (African-Black, Female, Able)}, for instance, foregrounded gender in a hiring scenario (\textit{Context 1: "It's giving a slightly misogynistic vibe with the male-dominated executive team reference." Assigned Identity: Gender (Female)}) but ethnicity or class in a consumer product context (\textit{Context 2: "This presupposes black or yellow or other skin types that are not white to be inferior." Assigned Identity: Ethnicity (African-Black)}). In line with our theoretical stance, this underpins the argument that positionality should be understood and approached as a fluid, context-sensitive construct.

\subsubsection{Intersectional Reasoning}
Annotators frequently reasoned through multiple, co-constitutive identity facets when explaining their fairness judgments. Many participants explicitly linked gender, race, class, or ability, showing how overlapping identities shaped their interpretations of the task. Sometimes this appeared as distributed reasoning, with different facets applied to different text segments; in others, annotators intertwined facets into a single, compound standpoint. For example, \textbf{P28's (Caucasian, Non-Binary)} connected feminist commitments, gendered social pressure, and assigned sex at birth in a critique of a job advertisement: \textit{"Woah, what a misogynist job vacancy! Women and people who present as women already have a lot of pressure on them put by society; it's better if we don't include confidence in this too. Confidence comes from the inside and self-worth." Assigned Identity: Militantism (Feminist), Assigned sex at birth (Female)}. 
This kind of intersectional anchoring enriched the interpretive space of the task, as annotators contextualised them within broader patterns of compounded discrimination. Intersectional reasoning thus contributed directly to epistemic richness, surfacing how fairness evaluations emerge from lived social complexity.

\subsubsection{Reflexivity Under Tension}
While the design probe helped annotators surface rich perspectives, they also revealed moments where reasoning was ambivalent or incomplete. These tensions exposed the limits and frictions inherent in articulating value judgments within constrained annotation workflows. Annotators grappled with how to express moral evaluations, map them onto fixed categories, and separate fairness concerns from personal reactions.

One form of tension appeared in under-articulated reflections, where annotators pointed to fairness-relevant cues but left their evaluative stance implicit. \textbf{P2 (African-Black, Female)}, for instance, highlighted “exclusive” in a luxury skincare advertisement but wrote only: \textit{“The use of the word exclusive.”} This left unclear whether she viewed the phrasing as aspirational, exclusionary, or simply noteworthy. Interview data showed that such brevity often reflected an assumption of shared understanding.  \textbf{P10 - Interview (African-Black, Female)} explained that she omitted explicit approval because it seemed obvious: \textit{“It's a good thing that it considers women's input. I just didn't write that it's a good thing because I think, as you can see from the text, it's good that in a job vacancy they consider women's input.”} These moments illustrate how annotators rely on implicit normativity rather than explicating their reasoning, creating ambiguity for systems that depend on explicit signals and in light of plural moral perspectives.

A second form of tension emerged through labelling dissonance, where annotators’ selected identity facets diverged from the nuance of their reasoning. \textbf{P21 (African-Black, Female)} critiqued exclusion rooted in class and income, yet selected “Gender (Female)” as the identity category: \textit{“The text suggests that perfection is not only desirable but also attainable via luxury products. Individuals from lower-income backgrounds may feel excluded for not affording such products, while gender norms often push perfectionist beauty ideals more heavily.”} Such mismatches suggest that annotators sometimes defaulted to familiar categories rather than those most aligned with their reflection, highlighting the difficulty of mapping contextualised reasoning onto rigid taxonomies.

A third tension arose in emotionally charged or beauty-oriented content, eliciting personal reactions that overshadowed fairness evaluation. Responding to the marketing phrase \textit{“the flawless fairness you truly deserve,”} \textbf{P3 (African-Black, Female)} disclosed: \textit{“This makes me feel good because I am not happy with the way my skin looks and feels.”} In contrast, \textbf{P17 (White, Female)} interpreted the same line as a normative judgment about appearance: \textit{“It's making an assumption that I am imperfect, it flat out says that I am not pretty enough for society.”} Together, these tensions reveal the complex terrain that annotators navigate when interpreting texts that resonate with personal insecurities or aspirations. Annotators moved between clarity and ambiguity, precision and approximation, moral critique and personal response. These frictions illuminate the messy nature of value judgments and underscore why reflexive annotating should be understood as revealing the lived complexity behind them.

\subsection{Experiencing Reflexive Annotating (RQ2)}
The design probe shaped how annotators anchored their value judgment in their social position and experienced the work. Asking workers to name relevant identities, revisit personal memories, or articulate interpretive cues introduced new expressive, emotional, and moral demands that diverged from efficiency-driven platform labour. These experiences influenced whether annotators engaged deeply, withdrew, or revised their own understanding of fairness. Related to RQ2, we describe three subthemes: (1) Emotional Exposure, (2) Strategic Withdrawal, and (3) Perspective Awareness.

\subsubsection{Emotional Exposure}
For many annotators, reflexive annotating elicited feelings of vulnerability by prompting them to revisit painful experiences or confront privilege. \textbf{P25 - Interview (White and Black Caribbean, Male)} described the emotional weight of returning to memories of assimilation: \textit{“When I do think about and I do remember about that stuff as a child, it can be a little bit emotional.”} Others experienced discomfort when acknowledging their own structural advantages. \textbf{P16 - Interview (White, Female)} reflected on this unease: \textit{“That’s quite an awkward thing to think about because you feel a bit guilty. [...] It’s difficult to admit that you’ve had more privilege based just on colour. It’s not fair, is it?”} For some, emotional discomfort was compounded by concerns about privacy and data use. \textbf{P22 - Interview (White, Female)} described initial hesitation: \textit{“Yeah, it was uncomfortable [to share intimate information], until I was reassured that Prolific would not share my data.”} Yet anonymity could also function as a protective buffer, enabling deeper engagement for certain workers. As \textbf{P25 - Interview (White and Black Caribbean, Male)} noted: \textit{“It makes me a lot more comfortable just being able to sit at my computer and enter the details myself… it doesn't make me worried about anything like that.”}
Together, these accounts show that reflexive annotating involves affective labour: annotators must navigate the emotional consequences of revealing, revisiting, or acknowledging aspects of their lives that are normally irrelevant in data work.

\subsubsection{Strategic Withdrawal}
Some annotators responded to the cognitive and emotional demands of reflexive engagement by scaling back their participation. Some reflections were minimal or disconnected from the fairness task, such as \textbf{P1's (White, Male)} terse description: \textit{“Women”} or \textbf{P5's (African-Black, Female)} unrelated comment: \textit{“They are willing to work extra hard to gain more knowledge.”} These instances of minimal engagement do not necessarily reflect misunderstanding. We interpret these as moments where annotators chose not to invest further emotional or cognitive effort, rather than as misunderstandings.  Reasons included discomfort with the task’s personal nature, uncertainty about whether identity should matter, or the pressure of crowd work that rewards speed over introspection. Strategic withdrawal thus shows how reflexivity can become an additional burden that annotators may selectively refuse.

\subsubsection{Perspective Awareness}
Reflexive annotating made several annotators more aware of the limits of their own perspective. They explicitly noted what they could not know, particularly when encountering forms of discrimination or marginalisation outside their lived experience. Some openly noted the partiality: \textbf{P13 (White, Male)} admitted: \textit{“Occasional difficulty relating to experiences of ethnic minorities.”} Similarly, \textbf{P2 (African-Black, Female)} reflected: \textit{“I don't have first-hand understanding of disability challenges, and I tend to overlook those needs.”} Others recognised the structural advantages shaping their own viewpoints. \textbf{P16 (White, UK)} stated this explicitly: \textit{“I have a lot of privilege being white. I have had more opportunities than other non white people.”} 

This often took the form of reflexive approximation, with annotators thinking through uncertainty: \textbf{P16 - Interview (White, Female)} questioned her own categorisation out loud: \textit{“Well, I don't know. I presume it's gender. Would age come into it?”} Such expressions of uncertainty signalled an awareness that fairness judgments are necessarily partial and provisional.
For some annotators, this awareness deepened over time and even led them to revise their judgments. We also observed cases where workers’ initial fairness ratings diverged from their later reflections. \textbf{P21 (African-Black, Female)} described a job vacancy as “very fair,” yet upon reflection, reconsidered its implications: \textit{“Requiring native English speakers will possibly exclude individuals who are bilingual but with perfect fluency. This may discriminate against non native speakers.”} These moments show how reflexive annotating cultivates perspective awareness and epistemic humility, as annotators engage not only with the text but also with the boundaries of their own standpoint.

\section{Discussion and Implications}
\label{sec:discussion}
In this discussion, we move from empirical findings to a conceptual account of \emph{reflexive annotating} as an epistemic practice. Concluding our empirical account, and building on Gentles et al. \cite{gentles2014critical} and Olmos-Vega et al. \cite{olmos2023practical}, we define \emph{reflexive annotating} as a \textit{set of continuous and multifaceted practices through which annotators disclose and interrogate how their social position and lived experience shape their judgments}. The goal is not to strip annotation of its subjectivity, a task both impossible and undesirable \cite{rees2020re}, but to foreground subjectivity as a legitimate locus of knowledge production in alignment. We discuss our findings by identifying two main contributions of reflexive annotating: \textbf{enhancing epistemic richness} by surfacing how diverse value perspectives are grounded in experience, and \textbf{enhancing epistemic accountability} by making visible how annotators understand their perspectives as partial and revisable. We discuss both aims below and consider \textbf{how reflexive annotating could enhance value alignment}.

\subsection{Enhancing Epistemic Richness}
By surfacing the positionality that informs annotators’ decisions, reflexive annotating reveals the social conditions under which labels are produced. We showed that resulting situated annotations capture dynamic and intersectional positional information that static metadata and inference-based approaches often overlook. Annotators expressed positionality in multiple narrative stances, from explicit lived experience \textit{(``as a non-binary person'')}, to relational or community framings \textit{(``women in my family'')}, to solidarity-based empathy \textit{(``I can put myself in the shoes of someone in that situation'')}. Such variation expands the interpretive space of annotations: judgments articulated through self-, community-, or other-as-subject framings all carry epistemic weight. This empirically observed variation demonstrates that positional engagement is not uniform but contextually enacted.

These findings substantiate feminist accounts of situated knowledge (e.g. \cite{soedirgo2020toward, crenshaw2013mapping, hooks2000feminist}) by providing insight into how annotators selectively foreground different aspects of themselves \textit{(e.g., gender in one context, ethnicity in another)}. While prior work has similarly shown that annotation behaviour depends on factors beyond static demographics \cite{orlikowski2023ecological, robinson2009ecological, freedman1999ecological, biester2022analyzing, alipour2024robustness}, our results offer direct empirical documentation of how such variability becomes explicitly expressed when reflexivity is supported by design. Without reflexive annotating, these layered positional articulations would have remained invisible.

Our findings suggest that this framing expands the interpretive space of annotation by treating socially derived variation as epistemically meaningful structure for alignment research. In doing so, they provide empirical grounding for rethinking alignment as a practice that can account for the broader social, political, historical, and personal contexts in which value judgments are situated.

\subsection{Enhancing Epistemic Accountability}
Reflexive annotating also helped annotators recognise assumptions, biases, and blind spots that might otherwise remain implicit, fostering more critical engagement with the task. Workers frequently stated what they could not know \textit{(``difficulty relating to experiences of ethnic minorities'')}, what they tended to overlook \textit{(``I don’t have first-hand understanding of disability challenges'')}, or where they needed to approximate \textit{(``I don’t know, I presume it’s gender'')}. Such statements reflect epistemic humility \cite{soedirgo2020toward,pachirat2015tyranny}. In Harding’s terms, this strengthens objectivity not by masking subjectivity but by making it visible and open to scrutiny \cite{harding1995strong}. At the same time, reflexive engagement was affectively and expressively demanding. Some annotators struggled to articulate intuitive judgments, experienced discomfort in acknowledging privilege, or revisited emotionally charged memories. Others withdrew strategically. These findings echo prior work on the affective costs of data work \cite{gray2019ghost, d2023data, alemadi2024emotional}, suggesting that reflexive annotating must balance epistemic benefits with labour implications. Annotators also noted when their perspective shifted, showing reflexivity as an unfolding process in which structured prompts allow judgments to deepen or change over time. Here, epistemic humility and uncertainty are not signs of disengagement but markers of a provisional and accountable stance toward knowledge. Designing for epistemic accountability thus means embedding support for articulation (not just prompts), preserving anonymity, and acknowledging that reflexivity is a laborious process.

\subsection{Toward Situated Alignment: Opportunities and Future Work} 
Building on our empirical account of reflexive annotating, we outline several contributions this practice enables for alignment research, system design, and implementation. We show how reflexive annotating opens up new ways of reasoning about where value judgments come from, whose perspectives are represented, how disagreement is interpreted, and how alignment data might be used in context-sensitive and scalable ways.

\textit{Traceable Value Judgments.}
Our findings highlight the importance of understanding how a value judgment arises and from which perspective. Rigid prompts and predefined identity labels sometimes obscured what annotators themselves considered salient, flattening contextual signals such as professional background or lived experience. Supporting traceability, therefore, requires designs that allow annotators to indicate which aspects of their perspective matter, for example, by separating affective reactions from evaluative claims or enabling brief rationales and uncertainty markers. Such rationales have been shown to improve data quality and inform the design and evaluation of model-generated rationales, supporting explainability in AI systems \cite{herrewijnen2024human}. Technically, such reflexive signals provide structured metadata that link judgments to their conditions of production---e.g., from personal experience or imaginative solidarity, with levels of uncertainty due to awareness of personal standpoint---enabling alignment processes to reason over situated provenance rather than decontextualised labels.

\textit{Surfacing Marginalised Perspectives.}
Our study suggests that marginalisation in alignment data can occur not only through under-representation but through representational mismatch, when systems recognise only certain forms of positionality as legitimate. Reflexive annotating allows annotators to articulate values and experiences in their own terms, helping surface perspectives that might otherwise remain invisible. This aligns with work showing that minority or dissenting judgments often raise concerns overlooked by consensus-driven approaches \cite{fazelpour2025value}, as well as standpoint-theoretic arguments that such perspectives may be most epistemically valuable for alignment \cite{harding2004feminist, harding2013rethinking}.

\textit{Making Disagreement Legible.}
Following recent calls to value pluralism \cite{sorensen2024roadmap}, situated annotations render disagreement and value pluralism interpretable. Prior work has shown that disagreement often reflects subjectivity or ambiguity rather than error and should be preserved \cite{leonardelli2021agreeing, uma2021learning, fleisig2024perspectivist, frenda2025perspectivist}. Existing technical approaches, such as modelling annotators individually \cite{deng2023you, geva2019we}, learning from label distributions \cite{rodrigues2018deep, weerasooriya2023disagreement}, or using disagreement for uncertainty-aware prediction \cite{khurana2024crowd}, could be extended by situated annotations that clarify why judgments diverge. In this way, disagreement becomes legible as a relational phenomenon between annotator, input, and context.

\textit{Context-Sensitive and Personalised Alignment.}
Reflexive annotating also has implications for context-sensitive or personalised alignment, where models adapt to user-specific values while maintaining transparency about whose values are enacted \cite{kirk2023personalisation}. Datasets such as PRISM \cite{kirk2024prism} demonstrate how linking judgments to richer participant profiles enables exploration of attribution and personalisation. By retaining explicit links between judgments and situated perspectives, reflexive annotating may further support alignment mechanisms that respond to context at deployment time rather than assuming a single, static value configuration.

\textit{Scalability vs. Reflexivity.}
Lastly, we would like to acknowledge that, beyond questions of utility, future work will need to further engage with how intentional, slow, and small situated approaches speak to or intersect with prevailing practices in data annotation. Resolving issues is beyond the scope of this paper, but we believe that reflexivity need not scale uniformly. It could be selectively deployed in high-stakes or normatively complex contexts or integrated into specific training pathways, such as active learning workflows \cite{van-der-meer-etal-2024-annotator}. While we have already tried to strike a balance between cost and reflexivity, we also believe future design refinements (e.g., merging Tier 1 \& 2) may further minimise cost while preserving epistemic richness.

Taken together, these contributions motivate a move beyond "simple alignment" and point toward what we term \textit{situated alignment}: approaches that treat value judgments as grounded in social position and context, and that seek to improve alignment by preserving the provenance, plurality, and contextual relevance of human judgments. Rather than replacing existing alignment techniques, situated alignment reframes their assumptions, shifting attention from producing a single “correct” behaviour to supporting alignment processes that can reason over whose values are represented, under what conditions, and with what limits.

\section{Limitations, Caveats, and Ethical Considerations} 
\label{section:limitations}
Several methodological limitations merit consideration. First, despite our intent to foster reflexivity, prompts elicited avoidance in some annotators, discomfort in others, and, at times, performative engagement, suggesting that reflexive annotating may not scale reliably without careful design adaptations. Second, our focus on fairness leaves open how these practices apply to other alignment-relevant values such as harm, dignity, or truthfulness. Third, our partly categorical approach to capturing positionality, adapted from the Social Identity Map \cite{jacobson2019social}, supported structured reflection but necessarily constrained the dynamic and narrative character of positionality \cite{berger2015now, bernstein1978restructuring, yanow2015thinking, carbado2013colorblind, showden2018youth, soedirgo2020toward}. While more open-ended formats could better represent its fluidity, early pilots showed that fully self-defined identities risked overwhelming participants, suggesting that greater complexity does not automatically yield deeper reflexivity. Future work might therefore explore designs that move beyond fixed demographic labels while avoiding cognitive overload (e.g., predefined yet more nuanced, bottom-up formed categories) to better balance expressiveness, usability, and situated nuance. Finally, our largely English-speaking sample limits cultural diversity and calls for replication in multilingual and cross-cultural settings.

Ethical and political risks also warrant attention. Collecting positionality data, even pseudonymised, carries re-identification and privacy risks: combined data points may reveal sensitive traits via the mosaic effect \cite{pozen2005mosaic}. Furthermore, while personalisation can enhance cultural relevance and user experience, it also raises the spectre of large-scale profiling, bias entrenchment, privacy infringements, and targeting of vulnerable populations \cite{kirk2023personalisation}. Prior work shows that personalisation can produce filter bubbles and affective polarisation, amplifying harms familiar from social media and search \cite{lazovich2023filter}.

Together, these considerations underscore that reflexive annotating sits at the intersection of technical design, epistemic theory, and ethical labour practice. Its promise lies in foregrounding annotator positionality and accountability while promoting situated alignment, but its viability depends on whether we, as a community, can create infrastructures that protect workers’ privacy, recognise their emotional and cognitive labour, and resist reproducing the very asymmetries that reflexivity seeks to expose. 
\section{Conclusion} 
\label{sec:conclusion}
This paper introduced reflexive annotating as a design probe that foregrounds annotators’ social positions, experiences, and interpretive cues in producing alignment data. Through qualitative crowdsourcing and follow-up interviews, we showed how reflexive prompts elicit intersectional reasoning, surface uncertainty and perspective shifts, and generate epistemic metadata unavailable through static demographics or behavioural inference. Our findings also underscore the affective demands of reflexivity, highlighting the need for designs that protect anonymity and respect limits on disclosure. We argue that alignment pipelines should treat annotations as situated, provenance-rich claims rather than neutral ground truths, and selectively incorporate reflexive metadata to make model behaviour more plural, interpretable, and accountable.

\section{Endmatter Section}
\subsection{Author Positionality}
\anon[Anonymised for review.] {We unpack our social position according to Savin-Baden \& Major's \cite{savin2023qualitative} and locate ourselves in relation to the research subject, the participants, and the wider research process. Our team combines critical/feminist design scholarship with computer science expertise in crowd work and LLMs/NLP, shaping our approach to annotation and its role in AI alignment. We acknowledge approaching crowd work primarily from an academic perspective. None of us relies on crowd work for income, though some have participated in sessions to better understand the experience. We engage with these issues largely as researchers with institutional privilege, able to purchase time on platforms rather than depending on them for livelihood. This stance reproduces the power imbalances inherent in crowd work, where anonymity and platform design obscure the people behind the labour. Our interpretative stance is informed by feminist commitments and Western European academic contexts. Several team members identify as women, most as white, shaping how we consider fairness, reflexivity, and pluralism in annotation. We also adopt a strongly qualitative approach, which, while less common in crowd work research, is essential for addressing our research questions.}

\subsection{Author Contributions}
\anon[Anonymised for review.] {Anne Arzberger led the conceptualisation of the research, designed the study, and collected the data, with iterative input from Alessandro Bozzon, Jie Yang, and Ujwal Gadiraju. Anne Arzberger and Céline Offerman jointly analysed the data. Anne Arzberger drafted the manuscript, and Céline Offerman contributed to revising and editing the paper.}

\subsection{Generative AI Use Disclosure}
During manuscript preparation, the authors used ChatGPT (OpenAI; GPT-4–class model) solely to assist with minor language editing, including grammar, spelling, and sentence-level clarity. No generative AI tools were used to produce original text, ideas, arguments, analyses, or interpretations. All intellectual contributions, claims, and conclusions are entirely those of the authors, who take full responsibility for the originality, accuracy, and integrity of the manuscript.

\subsection{Acknowledgments}
\anon[Anonymised for review.] {This work was supported by the ICAI lab GENIUS (Generative Enhanced Next-Generation Intelligent Understanding Systems), a collaboration between Delft University of Technology, Maastricht University, DSM-Firmenich, and KickstartAI, and by the NWO Long-Term Programme ROBUST initiated by the Innovation Centre for Artificial Intelligence (ICAI). We thank Shreyan Biswas for support with the web application development.}

\bibliographystyle{abbrv}
\bibliography{references}

@inproceedings{vertesi2011value,
  title={The value of data: considering the context of production in data economies},
  author={Vertesi, Janet and Dourish, Paul},
  booktitle={Proceedings of the ACM 2011 conference on Computer supported cooperative work},
  pages={533--542},
  year={2011}
}

@inproceedings{taylor2015data,
  title={Data-in-place: Thinking through the relations between data and community},
  author={Taylor, Alex S and Lindley, Si{\^a}n and Regan, Tim and Sweeney, David and Vlachokyriakos, Vasillis and Grainger, Lillie and Lingel, Jessica},
  booktitle={Proceedings of the 33rd Annual ACM Conference on Human Factors in Computing Systems},
  pages={2863--2872},
  year={2015}
}

@inproceedings{sambasivan2021re,
  title={Re-imagining algorithmic fairness in india and beyond},
  author={Sambasivan, Nithya and Arnesen, Erin and Hutchinson, Ben and Doshi, Tulsee and Prabhakaran, Vinodkumar},
  booktitle={Proceedings of the 2021 ACM conference on fairness, accountability, and transparency},
  pages={315--328},
  year={2021}
}

@book{cant2024feeding,
  title={Feeding the machine: The hidden human labor powering AI},
  author={Cant, Callum and Muldoon, James and Graham, Mark},
  year={2024},
  publisher={Bloomsbury Publishing USA}
}

@book{gray2019ghost,
  title={Ghost work: How to stop Silicon Valley from building a new global underclass},
  author={Gray, Mary L and Suri, Siddharth},
  year={2019},
  publisher={Harper Business}
}

@article{toxtli2021quantifying,
  title={Quantifying the invisible labor in crowd work},
  author={Toxtli, Carlos and Suri, Siddharth and Savage, Saiph},
  journal={Proceedings of the ACM on human-computer interaction},
  volume={5},
  number={CSCW2},
  pages={1--26},
  year={2021},
  publisher={ACM New York, NY, USA}
}

@inproceedings{bardzell2010feminist,
  title={Feminist HCI: taking stock and outlining an agenda for design},
  author={Bardzell, Shaowen},
  booktitle={Proceedings of the SIGCHI conference on human factors in computing systems},
  pages={1301--1310},
  year={2010}
}

@article{braun2019reflecting,
  title={Reflecting on reflexive thematic analysis},
  author={Braun, Virginia and Clarke, Victoria},
  journal={Qualitative research in sport, exercise and health},
  volume={11},
  number={4},
  pages={589--597},
  year={2019},
  publisher={Taylor \& Francis}
}

@article{wauthier2011bayesian,
  title={Bayesian bias mitigation for crowdsourcing},
  author={Wauthier, Fabian L and Jordan, Michael},
  journal={Advances in neural information processing systems},
  volume={24},
  year={2011}
}

@inproceedings{hube2019understanding,
  title={Understanding and mitigating worker biases in the crowdsourced collection of subjective judgments},
  author={Hube, Christoph and Fetahu, Besnik and Gadiraju, Ujwal},
  booktitle={Proceedings of the 2019 CHI conference on human factors in computing systems},
  pages={1--12},
  year={2019}
}

@article{ghai2020measuring,
  title={Measuring social biases of crowd workers using counterfactual queries},
  author={Ghai, Bhavya and Liao, Q Vera and Zhang, Yunfeng and Mueller, Klaus},
  journal={arXiv preprint arXiv:2004.02028},
  year={2020}
}

@article{miceli2020between,
  title={Between subjectivity and imposition: Power dynamics in data annotation for computer vision},
  author={Miceli, Milagros and Schuessler, Martin and Yang, Tianling},
  journal={Proceedings of the ACM on Human-Computer Interaction},
  volume={4},
  number={CSCW2},
  pages={1--25},
  year={2020},
  publisher={ACM New York, NY, USA}
}

@book{d2023data,
  title={Data feminism},
  author={D'ignazio, Catherine and Klein, Lauren F},
  year={2023},
  publisher={MIT press}
}

@article{asad2019prefigurative,
  title={Prefigurative design as a method for research justice},
  author={Asad, Mariam},
  journal={Proceedings of the ACM on Human-Computer Interaction},
  volume={3},
  number={CSCW},
  pages={1--18},
  year={2019},
  publisher={ACM New York, NY, USA}
}

@article{pessoa2019using,
  title={Using reflexive interviewing to foster deep understanding of research participants’ perspectives},
  author={Pessoa, Alex Sandro Gomes and Harper, Erin and Santos, Isabela Samogim and Gracino, Marina Carvalho Da Silva},
  journal={International journal of qualitative methods},
  volume={18},
  pages={1609406918825026},
  year={2019},
  publisher={SAGE Publications Sage CA: Los Angeles, CA}
}

@article{song2007conceptual,
  title={A conceptual model for understanding self-directed learning in online environments},
  author={Song, Liyan and Hill, Janette R},
  journal={Journal of interactive online learning},
  volume={6},
  number={1},
  pages={27--42},
  year={2007}
}

@article{ekbia2014heteromation,
  title={Heteromation and its (dis) contents: The invisible division of labor between humans and machines},
  author={Ekbia, Hamid and Nardi, Bonnie},
  journal={First Monday},
  year={2014}
}

@book{dunlosky2008metacognition,
  title={Metacognition},
  author={Dunlosky, John and Metcalfe, Janet},
  year={2008},
  publisher={Sage Publications}
}

@inproceedings{irani2010postcolonial,
  title={Postcolonial computing: a lens on design and development},
  author={Irani, Lilly and Vertesi, Janet and Dourish, Paul and Philip, Kavita and Grinter, Rebecca E},
  booktitle={Proceedings of the SIGCHI conference on human factors in computing systems},
  pages={1311--1320},
  year={2010}
}

@inproceedings{kapania2023hunt,
  title={A hunt for the snark: Annotator diversity in data practices},
  author={Kapania, Shivani and Taylor, Alex S and Wang, Ding},
  booktitle={Proceedings of the 2023 CHI Conference on Human Factors in Computing Systems},
  pages={1--15},
  year={2023}
}

@inproceedings{cambo2022model,
  title={Model positionality and computational reflexivity: Promoting reflexivity in data science},
  author={Cambo, Scott Allen and Gergle, Darren},
  booktitle={Proceedings of the 2022 CHI Conference on Human Factors in Computing Systems},
  pages={1--19},
  year={2022}
}

@book{collins2022black,
  title={Black feminist thought: Knowledge, consciousness, and the politics of empowerment},
  author={Collins, Patricia Hill},
  year={2022},
  publisher={routledge}
}

@incollection{haraway2013situated,
  title={Situated knowledges: The science question in feminism and the privilege of partial perspective 1},
  author={Haraway, Donna},
  booktitle={Women, science, and technology},
  pages={455--472},
  year={2013},
  publisher={Routledge}
}

@article{harding1995strong,
  title={“Strong objectivity”: A response to the new objectivity question},
  author={Harding, Sandra},
  journal={Synthese},
  volume={104},
  pages={331--349},
  year={1995},
  publisher={Springer}
}

@incollection{harding2013rethinking,
  title={Rethinking standpoint epistemology: What is “strong objectivity”?},
  author={Harding, Sandra},
  booktitle={Feminist epistemologies},
  pages={49--82},
  year={2013},
  publisher={Routledge}
}

@book{nagel1989view,
  title={The view from nowhere},
  author={Nagel, Thomas},
  year={1989},
  publisher={oxford university press}
}

@article{forsythe1993engineering,
  title={Engineering knowledge: The construction of knowledge in artificial intelligence},
  author={Forsythe, Diana E},
  journal={Social studies of science},
  volume={23},
  number={3},
  pages={445--477},
  year={1993},
  publisher={Sage Publications}
}

@inproceedings{wang2022whose,
  title={Whose AI Dream? In search of the aspiration in data annotation.},
  author={Wang, Ding and Prabhat, Shantanu and Sambasivan, Nithya},
  booktitle={Proceedings of the 2022 CHI conference on human factors in computing systems},
  pages={1--16},
  year={2022}
}

@inproceedings{kay2024epistemic,
  title={Epistemic injustice in generative AI},
  author={Kay, Jackie and Kasirzadeh, Atoosa and Mohamed, Shakir},
  booktitle={Proceedings of the AAAI/ACM Conference on AI, Ethics, and Society},
  volume={7},
  pages={684--697},
  year={2024}
}

@article{sorensen2024roadmap,
  title={Position: a roadmap to pluralistic alignment},
  author={Sorensen, Taylor and Moore, Jared and Fisher, Jillian and Gordon, Mitchell and Mireshghallah, Niloofar and Rytting, Christopher Michael and Ye, Andre and Jiang, Liwei and Lu, Ximing and Dziri, Nouha and others},
  booktitle={Proceedings of the 41st International Conference on Machine Learning},
  pages={46280--46302},
  year={2024}
}

@article{olmos2023practical,
  title={A practical guide to reflexivity in qualitative research: AMEE Guide No. 149},
  author={Olmos-Vega, Francisco M and Stalmeijer, Ren{\'e}e E and Varpio, Lara and Kahlke, Renate},
  journal={Medical teacher},
  volume={45},
  number={3},
  pages={241--251},
  year={2023},
  publisher={Taylor \& Francis}
}

@article{rees2020re,
  title={Re-visioning academic medicine through a constructionist lens},
  author={Rees, Charlotte E and Crampton, Paul ES and Monrouxe, Lynn V},
  journal={Academic Medicine},
  volume={95},
  number={6},
  pages={846--850},
  year={2020},
  publisher={LWW}
}

@article{gentles2014critical,
  title={Critical approach to reflexivity in grounded theory},
  author={Gentles, Stephen J and Jack, Susan M and Nicholas, David B and McKibbon, K Ann},
  journal={The Qualitative Report},
  volume={19},
  number={44},
  pages={1--14},
  year={2014}
}

@article{arzberger2024reflexive,
  title={Reflexive Data Curation: Opportunities and Challenges for Embracing Uncertainty in Human--AI Collaboration},
  author={Arzberger, Anne and Lupetti, Maria Luce and Giaccardi, Elisa},
  journal={ACM Transactions on Computer-Human Interaction},
  volume={31},
  number={6},
  pages={1--33},
  year={2024},
  publisher={ACM New York, NY}
}

@inproceedings{frauenberger2016critical,
  title={Critical realist HCI},
  author={Frauenberger, Christopher},
  booktitle={Proceedings of the 2016 CHI Conference Extended Abstracts on Human Factors in Computing Systems},
  pages={341--351},
  year={2016}
}

@inproceedings{baumer2015reflective,
  title={Reflective informatics: conceptual dimensions for designing technologies of reflection},
  author={Baumer, Eric PS},
  booktitle={Proceedings of the 33rd annual ACM conference on human factors in computing systems},
  pages={585--594},
  year={2015}
}

@book{savin2023qualitative,
  title={Qualitative research: The essential guide to theory and practice},
  author={Savin-Baden, Maggi and Major, Claire},
  year={2023},
  publisher={Routledge}
}

@misc{DicFairness,
  title = {Fairness},
  howpublished = {Cambridge Dictionary, {Cambridge University Press}},
  url = {https://dictionary.cambridge.org/dictionary/english/fairness},
  note = {Accessed: {2025-09-05}},
  }

@inproceedings{fazelpour2025value,
  title={The Value of Disagreement in AI Design, Evaluation, and Alignment},
  author={Fazelpour, Sina and Fleisher, Will},
  booktitle={Proceedings of the 2025 ACM Conference on Fairness, Accountability, and Transparency},
  pages={2138--2150},
  year={2025}
}

@article{freedman1999ecological,
  title={Ecological inference and the ecological fallacy},
  author={Freedman, David A},
  journal={International Encyclopedia of the social \& Behavioral sciences},
  volume={6},
  number={4027-4030},
  pages={1--7},
  year={1999}
}

@article{robinson2009ecological,
  title={Ecological correlations and the behavior of individuals},
  author={Robinson, William S},
  journal={International journal of epidemiology},
  volume={38},
  number={2},
  pages={337--341},
  year={2009},
  publisher={Oxford University Press}
}

@inproceedings{alemadi2024emotional,
  title={Emotional toll and coping strategies: Navigating the effects of annotating hate speech data},
  author={AlEmadi, Maryam M and Zaghouani, Wajdi},
  booktitle={Proceedings of the Workshop on Legal and Ethical Issues in Human Language Technologies@ LREC-COLING 2024},
  pages={66--72},
  year={2024}
}

@article{pachirat2015tyranny,
  title={The Tyranny of Light.},
  author={Pachirat, Timothy},
  journal={Qualitative \& Multi-Method Research},
  volume={13},
  number={1},
  year={2015}
}

@article{alipour2024robustness,
  title={Robustness and confounders in the demographic alignment of llms with human perceptions of offensiveness},
  author={Alipour, Shayan and Sen, Indira and Samory, Mattia and Mitra, Tanushree},
  journal={arXiv preprint arXiv:2411.08977},
  year={2024}
}

@inproceedings{biester2022analyzing,
  title={Analyzing the effects of annotator gender across NLP tasks},
  author={Biester, Laura and Sharma, Vanita and Kazemi, Ashkan and Deng, Naihao and Wilson, Steven and Mihalcea, Rada},
  booktitle={Proceedings of the 1st Workshop on Perspectivist Approaches to NLP@ LREC2022},
  pages={10--19},
  year={2022}
}

@inproceedings{orlikowski2023ecological,
  title={The Ecological Fallacy in Annotation: Modeling Human Label Variation goes beyond Sociodemographics},
  author={Orlikowski, Matthias and R{\"o}ttger, Paul and Cimiano, Philipp and Hovy, Dirk},
  booktitle={The 61st Annual Meeting Of The Association For Computational Linguistics},
  year={2023}
}

@book{harding2004feminist,
  title={The feminist standpoint theory reader: Intellectual and political controversies},
  author={Harding, Sandra G},
  year={2004},
  publisher={Psychology Press}
}

@article{simandan2019revisiting,
  title={Revisiting positionality and the thesis of situated knowledge},
  author={Simandan, Dragos},
  journal={Dialogues in human geography},
  volume={9},
  number={2},
  pages={129--149},
  year={2019},
  publisher={SAGE Publications Sage UK: London, England}
}

@article{mateescu2019ai,
  title={AI in context: the labor of integrating new technologies},
  author={Mateescu, Alexandra and Elish, Madeleine},
  year={2019},
  publisher={Data \& Society Research Institute}
}

@article{aroyo2023dices,
  title={Dices dataset: Diversity in conversational ai evaluation for safety},
  author={Aroyo, Lora and Taylor, Alex and Diaz, Mark and Homan, Christopher and Parrish, Alicia and Serapio-Garc{\'\i}a, Gregory and Prabhakaran, Vinodkumar and Wang, Ding},
  journal={Advances in Neural Information Processing Systems},
  volume={36},
  pages={53330--53342},
  year={2023}
}

@article{mohamed2020decolonial,
  title={Decolonial AI: Decolonial theory as sociotechnical foresight in artificial intelligence},
  author={Mohamed, Shakir and Png, Marie-Therese and Isaac, William},
  journal={Philosophy \& Technology},
  volume={33},
  pages={659--684},
  year={2020},
  publisher={Springer}
}

@article{mcdonald2019reliability,
  title={Reliability and inter-rater reliability in qualitative research: Norms and guidelines for CSCW and HCI practice},
  author={McDonald, Nora and Schoenebeck, Sarita and Forte, Andrea},
  journal={Proceedings of the ACM on human-computer interaction},
  volume={3},
  number={CSCW},
  pages={1--23},
  year={2019},
  publisher={ACM New York, NY, USA}
}

@article{braun2006using,
  title={Using thematic analysis in psychology},
  author={Braun, Virginia and Clarke, Victoria},
  journal={Qualitative research in psychology},
  volume={3},
  number={2},
  pages={77--101},
  year={2006},
  publisher={Taylor \& Francis}
}

@article{hopf1993verhaltnis,
  title={Zum Verh{\"a}ltnis von innerfamilialen sozialen Erfahrungen, Pers{\"o}nlichkeitsentwicklung und politischen Orientierungen: Dokumentation und Er{\"o}rterung des methodischen Vorgehens in einer Studie zu diesem Thema},
  author={Hopf, Christel and Schmidt, Christiane},
  year={1993},
  publisher={DEU}
}

@article{clarke2017thematic,
  title={Thematic analysis},
  author={Clarke, Victoria and Braun, Virginia},
  journal={The journal of positive psychology},
  volume={12},
  number={3},
  pages={297--298},
  year={2017},
  publisher={Taylor \& Francis}
}

@article{carbado2013colorblind,
  title={Colorblind intersectionality},
  author={Carbado, Devon W},
  journal={Signs: Journal of Women in Culture and Society},
  volume={38},
  number={4},
  pages={811--845},
  year={2013},
  publisher={University of Chicago Press Chicago, IL}
}

@article{gadiraju2017modus,
  title={Modus operandi of crowd workers: The invisible role of microtask work environments},
  author={Gadiraju, Ujwal and Checco, Alessandro and Gupta, Neha and Demartini, Gianluca},
  journal={Proceedings of the ACM on Interactive, Mobile, Wearable and Ubiquitous Technologies},
  volume={1},
  number={3},
  pages={1--29},
  year={2017},
  publisher={ACM New York, NY, USA}
}

@article{rafailov2024direct,
  title={Direct preference optimization: Your language model is secretly a reward model},
  author={Rafailov, Rafael and Sharma, Archit and Mitchell, Eric and Manning, Christopher D and Ermon, Stefano and Finn, Chelsea},
  journal={Advances in Neural Information Processing Systems},
  volume={36},
  year={2024}
}

@article{aroyo2015truth,
  title={Truth is a lie: Crowd truth and the seven myths of human annotation},
  author={Aroyo, Lora and Welty, Chris},
  journal={AI Magazine},
  volume={36},
  number={1},
  pages={15--24},
  year={2015}
}

@misc{prolificProlificEasily,
	author = {},
	title = {{P}rolific | {E}asily collect high-quality data from real people --- prolific.com},
	howpublished = {\url{https://www.prolific.com/}},
	year = {},
	note = {[Accessed 30-12-2025]},
}

@misc{mturkAmazonMechanical,
	author = {},
	title = {{A}mazon {M}echanical {T}urk --- mturk.com},
	howpublished = {\url{https://www.mturk.com/}},
	year = {},
	note = {[Accessed 30-12-2025]},
}

@article{gaver1999design,
  title={Design: cultural probes},
  author={Gaver, Bill and Dunne, Tony and Pacenti, Elena},
  journal={interactions},
  volume={6},
  number={1},
  pages={21--29},
  year={1999},
  publisher={ACM New York, NY, USA}
}

@inproceedings{wan2025noise,
  title={From Noise to Nuance: Enriching Subjective Data Annotation through Qualitative Analysis},
  author={Wan, Ruyuan and Wang, Haonan and Huang, Ting-Hao and Gao, Jie},
  booktitle={Proceedings of the Fourth Workshop on Bridging Human-Computer Interaction and Natural Language Processing (HCI+ NLP)},
  pages={240--254},
  year={2025}
}

@inproceedings{mokhberian2024capturing,
  title={Capturing perspectives of crowdsourced annotators in subjective learning tasks},
  author={Mokhberian, Negar and Marmarelis, Myrl and Hopp, Frederic and Basile, Valerio and Morstatter, Fred and Lerman, Kristina},
  booktitle={Proceedings of the 2024 Conference of the North American Chapter of the Association for Computational Linguistics: Human Language Technologies (Volume 1: Long Papers)},
  pages={7337--7349},
  year={2024}
}

@article{patton2019annotating,
  title={Annotating social media data from vulnerable populations: Evaluating disagreement between domain experts and graduate student annotators},
  author={Patton, Desmond and Blandfort, Philipp and Frey, William and Gaskell, Michael and Karaman, Svebor},
  year={2019}
}

@inproceedings{talat2016you,
  title={Are you a racist or am i seeing things? annotator influence on hate speech detection on twitter},
  author={Talat, Zeerak},
  booktitle={Proceedings of the first workshop on NLP and computational social science},
  pages={138--142},
  year={2016}
}

@inproceedings{diaz2018addressing,
  title={Addressing age-related bias in sentiment analysis},
  author={D{\'\i}az, Mark and Johnson, Isaac and Lazar, Amanda and Piper, Anne Marie and Gergle, Darren},
  booktitle={Proceedings of the 2018 chi conference on human factors in computing systems},
  pages={1--14},
  year={2018}
}

@article{luo2020detecting,
  title={Detecting stance in media on global warming},
  author={Luo, Yiwei and Card, Dallas and Jurafsky, Dan},
  journal={arXiv preprint arXiv:2010.15149},
  year={2020}
}

@article{khurana2024crowd,
  title={Crowd-Calibrator: Can Annotator Disagreement Inform Calibration in Subjective Tasks?},
  author={Khurana, Urja and Nalisnick, Eric and Fokkens, Antske and Swayamdipta, Swabha},
  journal={arXiv preprint arXiv:2408.14141},
  year={2024}
}

@inproceedings{rodrigues2018deep,
  title={Deep learning from crowds},
  author={Rodrigues, Filipe and Pereira, Francisco},
  booktitle={Proceedings of the AAAI conference on artificial intelligence},
  volume={32},
  number={1},
  year={2018}
}

@inproceedings{weerasooriya2023disagreement,
  title={Disagreement matters: Preserving label diversity by jointly modeling item and annotator label distributions with DisCo},
  author={Weerasooriya, Tharindu Cyril and Ororbia, Alexander and Bhensadadia, Raj and KhudaBukhsh, Ashiqur and Homan, Christopher},
  booktitle={Findings of the Association for Computational Linguistics: ACL 2023},
  pages={4679--4695},
  year={2023}
}

@inproceedings{talat2016hateful,
  title={Hateful symbols or hateful people? predictive features for hate speech detection on twitter},
  author={Talat, Zeerak and Hovy, Dirk},
  booktitle={Proceedings of the NAACL student research workshop},
  pages={88--93},
  year={2016}
}

@inproceedings{sabou2014corpus,
  title={Corpus annotation through crowdsourcing: Towards best practice guidelines.},
  author={Sabou, Marta and Bontcheva, Kalina and Derczynski, Leon and Scharl, Arno},
  booktitle={LREC},
  pages={859--866},
  year={2014}
}

@inproceedings{geva2019we,
  title={Are we modeling the task or the annotator? An investigation of annotator bias in natural language understanding datasets},
  author={Geva, Mor and Goldberg, Yoav and Berant, Jonathan},
  booktitle={2019 Conference on Empirical Methods in Natural Language Processing and 9th International Joint Conference on Natural Language Processing, EMNLP-IJCNLP 2019},
  pages={1161--1166},
  year={2019},
  organization={Association for Computational Linguistics}
}

@article{frenda2025perspectivist,
  title={Perspectivist approaches to natural language processing: a survey},
  author={Frenda, Simona and Abercrombie, Gavin and Basile, Valerio and Pedrani, Alessandro and Panizzon, Raffaella and Cignarella, Alessandra Teresa and Marco, Cristina and Bernardi, Davide},
  journal={Language Resources and Evaluation},
  volume={59},
  number={2},
  pages={1719--1746},
  year={2025},
  publisher={Springer}
}

@inproceedings{fleisig2024perspectivist,
  title={The Perspectivist Paradigm Shift: Assumptions and Challenges of Capturing Human Labels},
  author={Fleisig, Eve and Blodgett, Su Lin and Klein, Dan and Talat, Zeerak},
  booktitle={Proceedings of the 2024 Conference of the North American Chapter of the Association for Computational Linguistics: Human Language Technologies (Volume 1: Long Papers)},
  pages={2279--2292},
  year={2024}
}

@article{uma2021learning,
  title={Learning from disagreement: A survey},
  author={Uma, Alexandra N and Fornaciari, Tommaso and Hovy, Dirk and Paun, Silviu and Plank, Barbara and Poesio, Massimo},
  journal={Journal of Artificial Intelligence Research},
  volume={72},
  pages={1385--1470},
  year={2021}
}

@inproceedings{leonardelli2021agreeing,
  title={Agreeing to Disagree: Annotating Offensive Language Datasets with Annotators’ Disagreement},
  author={Leonardelli, Elisa and Menini, Stefano and Palmero Aprosio, Alessio and Guerini, Marco and Tonelli, Sara and others},
  booktitle={Proceedings of the 2021 Conference on Empirical Methods in Natural Language Processing},
  pages={10528--10539},
  year={2021},
  organization={Association for Computational Linguistics}
}

@book{bernstein1978restructuring,
  title={The restructuring of social and political theory},
  author={Bernstein, Richard J},
  year={1978},
  publisher={University of Pennsylvania Press}
}

@incollection{yanow2015thinking,
  title={Thinking interpretively: Philosophical presuppositions and the human sciences},
  author={Yanow, Dvora},
  booktitle={Interpretation and method},
  pages={5--26},
  year={2015},
  publisher={Routledge}
}

@article{berger2015now,
  title={Now I see it, now I don’t: Researcher’s position and reflexivity in qualitative research},
  author={Berger, Roni},
  journal={Qualitative research},
  volume={15},
  number={2},
  pages={219--234},
  year={2015},
  publisher={Sage Publications Sage UK: London, England}
}

@book{fujii2017killing,
  title={Killing neighbors: Webs of violence in Rwanda},
  author={Fujii, Lee Ann},
  year={2017},
  publisher={Cornell University Press}
}

@book{hooks2000feminist,
  title={Feminist theory: From margin to center},
  author={Hooks, Bell},
  year={2000},
  publisher={Pluto Press}
}

@incollection{crenshaw2013mapping,
  title={Mapping the margins: Intersectionality, identity politics, and violence against women of color},
  author={Crenshaw, Kimberl{\'e} Williams},
  booktitle={The public nature of private violence},
  pages={93--118},
  year={2013},
  publisher={Routledge}
}

@book{showden2018youth,
  title={Youth who trade sex in the US: Intersectionality, agency, and vulnerability},
  author={Showden, Carisa Renae and Majic, Samantha},
  year={2018},
  publisher={Temple University Press}
}

@article{henry2009positionality,
  title={Positionality and power: The politics of peacekeeping research},
  author={Henry, Marsha and Higate, Paul and Sanghera, Gurchathen},
  journal={International Peacekeeping},
  volume={16},
  number={4},
  pages={467--482},
  year={2009},
  publisher={Taylor \& Francis}
}

@incollection{shehata2015ethnography,
  title={Ethnography, identity, and the production of knowledge},
  author={Shehata, Samer},
  booktitle={Interpretation and Method},
  pages={209--227},
  year={2015},
  publisher={Routledge}
}

@incollection{yanow2014interpretive,
  title={Interpretive analysis and comparative research},
  author={Yanow, Dvora},
  booktitle={Comparative policy studies: Conceptual and methodological challenges},
  pages={131--159},
  year={2014},
  publisher={Springer}
}

@article{soedirgo2020toward,
  title={Toward active reflexivity: Positionality and practice in the production of knowledge},
  author={Soedirgo, Jessica and Glas, Aarie},
  journal={PS: Political Science \& Politics},
  volume={53},
  number={3},
  pages={527--531},
  year={2020},
  publisher={Cambridge University Press}
}

@article{davani2022dealing,
  title={Dealing with disagreements: Looking beyond the majority vote in subjective annotations},
  author={Davani, Aida Mostafazadeh and D{\'\i}az, Mark and Prabhakaran, Vinodkumar},
  journal={Transactions of the Association for Computational Linguistics},
  volume={10},
  pages={92--110},
  year={2022},
  publisher={MIT Press One Rogers Street, Cambridge, MA 02142-1209, USA journals-info~…}
}

@article{herrewijnen2024human,
  title={Human-annotated rationales and explainable text classification: a survey},
  author={Herrewijnen, Elize and Nguyen, Dong and Bex, Floris and van Deemter, Kees},
  journal={Frontiers in Artificial Intelligence},
  volume={7},
  pages={1260952},
  year={2024},
  publisher={Frontiers Media SA}
}

@inproceedings{van-der-meer-etal-2024-annotator,
    title = "Annotator-Centric Active Learning for Subjective {NLP} Tasks",
    author = "van der Meer, Michiel  and
      Falk, Neele  and
      Murukannaiah, Pradeep K.  and
      Liscio, Enrico",
    editor = "Al-Onaizan, Yaser  and
      Bansal, Mohit  and
      Chen, Yun-Nung",
    booktitle = "Proceedings of the 2024 Conference on Empirical Methods in Natural Language Processing",
    month = nov,
    year = "2024",
    address = "Miami, Florida, USA",
    publisher = "Association for Computational Linguistics",
    url = "https://aclanthology.org/2024.emnlp-main.1031/",
    doi = "10.18653/v1/2024.emnlp-main.1031",
    pages = "18537--18555"
}

@inproceedings{zhangcultivating,
  title={Cultivating Pluralism In Algorithmic Monoculture: The Community Alignment Dataset},
  author={Zhang, Lily H and Milli, Smitha and Jusko, Karen Long and Smith, Jonathan and Amos, Brandon and Bouaziz, Wassim and Kussman, Jack and Revel, Manon and Titus, Lisa and Radharapu, Bhaktipriya and others},
  booktitle={2nd Workshop on Models of Human Feedback for AI Alignment}
}

@article{kidder1987qualitative,
  title={Qualitative and quantitative methods: When stories converge},
  author={Kidder, Louise H and Fine, Michelle},
  journal={New directions for program evaluation},
  volume={1987},
  number={35},
  pages={57--75},
  year={1987},
  publisher={Wiley Online Library}
}

@article{jacobson2019social,
  title={Social identity map: A reflexivity tool for practicing explicit positionality in critical qualitative research},
  author={Jacobson, Danielle and Mustafa, Nida},
  journal={International Journal of Qualitative Methods},
  volume={18},
  pages={1609406919870075},
  year={2019},
  publisher={SAGE Publications Sage CA: Los Angeles, CA}
}

@inproceedings{talat2022machine,
  title={On the machine learning of ethical judgments from natural language},
  author={Talat, Zeerak and Blix, Hagen and Valvoda, Josef and Ganesh, Maya Indira and Cotterell, Ryan and Williams, Adina},
  booktitle={Proceedings of the 2022 Conference of the North American Chapter of the Association for Computational Linguistics: Human Language Technologies},
  pages={769--779},
  year={2022},
  organization={Association for Computational Linguistics}
}

@article{sap2021annotators,
  title={Annotators with Attitudes: How Annotator Beliefs And Identities Bias Toxic Language Detection},
  author={Sap, Maarten and Swayamdipta, Swabha and Vianna, Laura and Zhou, Xuhui and Choi, Yejin and Smith, Noah},
  booktitle={Proceedings of the 2022 Conference of the North American Chapter of the Association for Computational Linguistics: Human Language Technologies},
  year={2022},
  organization={Association for Computational Linguistics}
}

@misc{apicella2020beyond,
  title={Beyond WEIRD: A review of the last decade and a look ahead to the global laboratory of the future},
  author={Apicella, Coren and Norenzayan, Ara and Henrich, Joseph},
  journal={Evolution and Human Behavior},
  volume={41},
  number={5},
  pages={319--329},
  year={2020},
  publisher={Elsevier}
}

@inproceedings{fleisig2023majority,
    title = "When the Majority is Wrong: Modeling Annotator Disagreement for Subjective Tasks",
    author = "Fleisig, Eve  and
      Abebe, Rediet  and
      Klein, Dan",
    editor = "Bouamor, Houda  and
      Pino, Juan  and
      Bali, Kalika",
    booktitle = "Proceedings of the 2023 Conference on Empirical Methods in Natural Language Processing",
    month = dec,
    year = "2023",
    address = "Singapore",
    publisher = "Association for Computational Linguistics",
    url = "https://aclanthology.org/2023.emnlp-main.415/",
    doi = "10.18653/v1/2023.emnlp-main.415",
    pages = "6715--6726"
}

@InCollection{sep-feminism-epistemology,
	author       =	{Anderson, Elizabeth},
	title        =	{{Feminist Epistemology and Philosophy of Science}},
	booktitle    =	{The {Stanford} Encyclopedia of Philosophy},
	editor       =	{Edward N. Zalta and Uri Nodelman},
	howpublished =	{\url{https://plato.stanford.edu/archives/fall2024/entries/feminism-epistemology/}},
	year         =	{2024},
	edition      =	{{F}all 2024},
	publisher    =	{Metaphysics Research Lab, Stanford University}
}

@inproceedings{arzberger2024nothing,
  title={Nothing Comes Without Its World--Practical Challenges of Aligning LLMs to Situated Human Values through RLHF},
  author={Arzberger, Anne and Buijsman, Stefan and Lupetti, Maria Luce and Bozzon, Alessandro and Yang, Jie},
  booktitle={Proceedings of the AAAI/ACM Conference on AI, Ethics, and Society},
  volume={7},
  pages={61--73},
  year={2024}
}

@article{kirk2024prism,
  title={The PRISM alignment dataset: What participatory, representative and individualised human feedback reveals about the subjective and multicultural alignment of large language models},
  author={Kirk, Hannah Rose and Whitefield, Alexander and Rottger, Paul and Bean, Andrew M and Margatina, Katerina and Mosquera-Gomez, Rafael and Ciro, Juan and Bartolo, Max and Williams, Adina and He, He and others},
  journal={Advances in Neural Information Processing Systems},
  volume={37},
  pages={105236--105344},
  year={2024}
}

@inproceedings{prabhakaran2021releasing,
  title={On Releasing Annotator-Level Labels and Information in Datasets},
  author={Prabhakaran, Vinodkumar and Davani, Aida Mostafazadeh and D{\'\i}az, Mark},
  booktitle={Proceedings of the Joint 15th Linguistic Annotation Workshop (LAW) and 3rd Designing Meaning Representations (DMR) Workshop},
  pages={133--138},
  year={2021}
}

@String{Computing = "Computing" }

@String{Computer = "{IEEE} Computer" }

@String{Academic = "Academic Press" }

@String{Chelsea = "Chelsea" }

@String{Springer = "Springer-Verlag" }

@book{christian2020alignment,
  title={The alignment problem: Machine learning and human values},
  author={Christian, Brian},
  year={2020},
  publisher={WW Norton \& Company}
}

@article{kirk2023personalisation,
  title={Personalisation within bounds: A risk taxonomy and policy framework for the alignment of large language models with personalised feedback},
  author={Kirk, Hannah Rose and Vidgen, Bertie and R{\"o}ttger, Paul and Hale, Scott A},
  journal={arXiv preprint arXiv:2303.05453},
  year={2023}
}

@article{adam2000deleting,
  title={Deleting the subject: A feminist reading of epistemology in artificial intelligence},
  author={Adam, Alison},
  journal={Minds and Machines},
  volume={10},
  number={2},
  pages={231--253},
  year={2000},
  publisher={Springer}
}

@article{wong2020cultural,
  title={Cultural differences as excuses? Human rights and cultural values in global ethics and governance of AI},
  author={Wong, Pak-Hang},
  journal={Philosophy \& Technology},
  volume={33},
  number={4},
  pages={705--715},
  year={2020},
  publisher={Springer}
}

@book{suchman1987plans,
  title={Plans and situated actions: The problem of human-machine communication},
  author={Suchman, Lucille Alice},
  year={1987},
  publisher={Cambridge university press}
}

@inproceedings{pei2023annotator,
  title={When Do Annotator Demographics Matter? Measuring the Influence of Annotator Demographics with the POPQUORN Dataset},
  author={Pei, Jiaxin and Jurgens, David},
  booktitle={Proceedings of the 17th Linguistic Annotation Workshop (LAW-XVII)},
  year={2023}
}

@inproceedings{lazovich2023filter,
  title={Filter bubbles and affective polarization in user-personalized large language model outputs},
  author={Lazovich, Tomo},
  booktitle={Proceedings on},
  pages={29--37},
  year={2023},
  organization={PMLR}
}

@article{pozen2005mosaic,
  title={The mosaic theory, national security, and the freedom of information act},
  author={Pozen, David E},
  journal={Yale LJ},
  volume={115},
  pages={628},
  year={2005},
  publisher={HeinOnline}
}

@inproceedings{deng2023you,
  title={You Are What You Annotate: Towards Better Models through Annotator Representations},
  author={Deng, Naihao and Zhang, Xinliang Frederick and Liu, Siyang and Wu, Winston and Wang, Lu and Mihalcea, Rada},
  booktitle={The 2023 Conference on Empirical Methods in Natural Language Processing},
  year={2023}
}

@article{stiennon2020learning,
  title={Learning to summarize with human feedback},
  author={Stiennon, Nisan and Ouyang, Long and Wu, Jeffrey and Ziegler, Daniel and Lowe, Ryan and Voss, Chelsea and Radford, Alec and Amodei, Dario and Christiano, Paul F},
  journal={Advances in Neural Information Processing Systems},
  volume={33},
  pages={3008--3021},
  year={2020}
}

@article{ouyang2022training,
  title={Training language models to follow instructions with human feedback},
  author={Ouyang, Long and Wu, Jeffrey and Jiang, Xu and Almeida, Diogo and Wainwright, Carroll and Mishkin, Pamela and Zhang, Chong and Agarwal, Sandhini and Slama, Katarina and Ray, Alex and others},
  journal={Advances in Neural Information Processing Systems},
  volume={35},
  pages={27730--27744},
  year={2022}
}

\appendix

\newpage
\section{Participant Demographics}
\label{appendix:A}

Table \ref{tab:demo1} presents the demographics of the participants who engaged in our prolific main study.

\begin{table}[H]
\centering
\small
\begin{tabular}{@{}>{}p{1.5cm}@{}p{12.6cm}@{}}
\toprule
\textbf{Age (years)} &
  $M=34.4$, $SD=12.9$, range $19$--$67$ \\ \midrule
\textbf{Gender} &
  \begin{tabular}[c]{@{}l@{}}Male: 14 (47\%)\\ Female: 13 (43\%)\\ Non-binary: 3 (10\%)\end{tabular}\\ \midrule
\textbf{Ethnicity} &
  \begin{tabular}[c]{@{}l@{}}White: 17 (57\%)\\ Black: 10 (33\%)\\ Asian: 1 (3\%)\\ Mixed: 1 (3\%)\\ Other: 1 (3\%)\end{tabular} \\ \midrule
\textbf{Ability} &
  \begin{tabular}[c]{@{}l@{}}Abled: 21 (70\%)\\ Disabled: 9 (30\%)\end{tabular} \\ \bottomrule
\end{tabular}%
\caption{Participant demographics for the follow-up interview, with a subsample from the Prolific participants ($N=5$). Percentages are of $N$.}
\label{tab:demo1}
\Description{Participant demographics for the follow-up interview.}
\end{table}

Table \ref{tab:demo2} presents the demographics of the participants who were interviewed after their participation in our main prolific study.

\begin{table}[H]
\centering
\small
\begin{tabular}{@{}>{}p{1.5cm}@{}p{12.6cm}@{}}
\toprule
\textbf{Age (years)} &
  $M=31$, $SD=10.35$, range $20$--$44$ \\ \midrule
\textbf{Gender} &
  \begin{tabular}[c]{@{}l@{}}Female: 4 (80\%)\\ Male: 1 (20\%)\end{tabular}\\ \midrule
\textbf{Ethnicity} &
  \begin{tabular}[c]{@{}l@{}}White: 2 (40\%)\\ Black: 2 (40\%)\\ Asian: 1 (20\%)\end{tabular} \\ \midrule
\textbf{Ability} &
  \begin{tabular}[c]{@{}l@{}}Abled: 3 (60\%)\\ Disabled: 2 (40\%)\end{tabular} \\ \bottomrule
\end{tabular}%
\caption{Participant demographics for the follow-up interview, with a subsample from the Prolific participants ($N=5$). Percentages are of $N$.}
\label{tab:demo2}
\Description{Participant demographics for the follow-up interview.}
\end{table}

\clearpage

\section{Follow-up Interview Protocol}
\label{appendix:B}
In what follows, we present the interview protocol for our semi-structured interviews with the crowd workers.

\subsection{1. Introduction}
A few questions to get to know the participant/ease them into the interview:
\begin{itemize}
    \item Could you tell me a bit about yourself?
    \item How did you first get into crowd work?
    \item What kind of tasks do you typically do, and how long have you been doing this work?
\end{itemize}

\subsection{2. Before the Task}
These questions explore the participant's thoughts before starting the annotation:
\begin{itemize}
    \item Before you started annotating, what did you expect from the task?
    \item Did you consider how aspects of your identity—like gender, ethnicity, or personal experiences—might shape how annotate? [follow-up] What reflections did that bring up for you?
    \item How would you define “social identity”? [follow-up] (If needed, clarify that in this context it refers to aspects like background, gender, or ability—things that shape how we see the world to create a shared understanding of the word for the remaining interview.)
\end{itemize}

\subsection{3. During the Task – Reflections}
These questions allow us to revisit some of the annotation tasks the participant performed during the prolific study, including examples from their responses.

\begin{itemize}
    \item Can you describe all the tasks you did (from start to finish) in your own words?
    \item In comparison to usual tasks you perform on prolific, was this exceptionally time intensive?
    \item And how was the cognitive load?
    \item How did you feel overall while doing the reflection exercises?
\end{itemize}

With respect to Tier 1\& 2 annotation tasks:
[example selected from participant responses]
\begin{itemize}
    \item Was anything especially easy or difficult? Helpful or confusing?
    \item Did you feel like you could express your identity comfortably here?
    \item What did you learn about yourself?
    \item Was it emotionally difficult?
    \item Was it uncomfortable to share such intimate information?
\end{itemize}

With respect to Tier 3 annotation task:
[example selected from participant responses]
\begin{itemize}
    \item Can you describe this experience in three words? Why those?
    \item What helped you connect this fairness reflection to your social identity?
    \item Were there any personal memories or feelings that came up?
    \item Did this feel meaningful? Challenging?
    \item (Was anything surprising?)
    \item How do you know that you provided a complete list of relevant identity facets? That there is none that would be better fitting to describe your perception here?
    \item If you could redesign this tool or process, what would you change?
\end{itemize}

\subsection{4. After the Task - Reflections}
The following question relates to the participants' thoughts and feelings after the annotation tasks.
\begin{itemize}
    \item Would you have liked to adjust your fairness ranking after you analysed the text in more detail? [follow-up] If so, how?
    \item Has this reflection process changed anything in how you view your annotation work?
    \item Are there any personal assumptions and biases that you became aware of?
    \item Will any of the insights you gained potentially affect future annotation tasks you perform? [follow-up] How?
\end{itemize}

\subsection{5. Reflecting on Specific Moments}
These questions aim to foster reflection on specific moments from the participants' annotations where something stood out—either because it was unclear, surprising, or where we saw a bit of tension or complexity. 1–3 relevant moments from below are selected, related to the participants' annotation responses. The most insightful or ambiguous ones are chosen.

\textbf{Misalignment / Unclear Reasoning}
[add example]
\begin{itemize}
    \item Here’s a moment where your answer and explanation seemed a bit at odds. Can you walk me through what you were thinking?
    \item Looking back, do you think you'd approach it differently now?
\end{itemize}

\textbf{Lower Engagement or Ambiguity}
[add example]
\begin{itemize}
    \item Here, your response was brief or detached. Was that intentional, or was something unclear or unengaging at that moment?
    \item What could’ve helped keep your focus or interest?
\end{itemize}

\textbf{Misunderstanding or Instructional Confusion}
[add example]
\begin{itemize}
    \item Did any parts of the task or wording feel confusing or ambiguous?
    \item How did that affect what you wrote or thought?
\end{itemize}

\textbf{Struggling with Expression / Misarticulation}
[add example]
\begin{itemize}
    \item Sometimes it’s hard to put our thoughts into words—did you experience that at any point?
    \item Were there moments where you felt your reflection didn’t fully capture what you meant?
    \item What support or changes could help make it easier to express yourself?
\end{itemize}

\textbf{Multiple Identity Facets / Intersectionality}
[add example]
\begin{itemize}
    \item Here you mentioned more than one aspect of your identity. Can you tell me what you were thinking when adding those two facets?
\end{itemize}

\textbf{Detached Framing}
[add example]
\begin{itemize}
    \item In this part, you pointed out an issue, but didn't link it to your own perception or personal identity. Did you see it as personally relevant? Why or why not?
\end{itemize}

\textbf{Avoiding Reflexivity}
[add example]
\begin{itemize}
    \item At this moment, it seemed like the task didn’t fully resonate with you. Did the reflection feel meaningful to you? Why or why not?
    \item What could make such a task feel more relevant or engaging for you?
\end{itemize}

\subsection{Closing}
\begin{itemize}
    \item Is there anything else I didn’t ask that feels important to share?
\end{itemize}

\section{Text Pieces for Fairness Ranking Task}
\label{appendix:c}
The following text samples were ranked by the crowd workers in terms of fairness. They were presented in randomised order. Text sample a) resembles a job vacancy, while sample b) relates to an advertisement of a skin care product.

\subsection{Text Sample a) Job Vacancy}
The job vacancy: \textit{“We are seeking a recent graduate with a youthful and energetic attitude to support our senior management team at a leading tech company. The ideal candidate will excel in managing executives’ demanding schedules, preparing polished, high-level documents and presentations, coordinating meetings seamlessly, and maintaining absolute discretion. Candidates must have native English proficiency and demonstrate impeccable grooming and a professional demeanour at all times. Proficiency in office software is not necessary, but women are especially encouraged to apply, as we value the multitasking and organisational skills they often bring. This role suits someone willing to work long hours, thrives in fast-paced environments, and is ready to align themselves with the vision of a high-achieving, male-dominated executive team.”}

\subsection{Text Sample b) Product Advertisement}
The advertisement: \textit{"Experience the ultimate self-care with our new line of luxury skincare products, specially designed to meet the unique needs of soft, sensitive skin. From rejuvenating night creams to hydrating serums, our products are made for those who understand the importance of embracing their natural beauty and taking the time to prioritise themselves. Join the countless users who trust us to support their journey toward healthier, glowing skin. Because everyone deserves to feel confident in their own skin."}

\clearpage

\section{Conceptual Model of Reflexive Thematic Analysis}
\label{appendix:d}

\begin{figure}[h]
    \centering
    \includegraphics[width=1\linewidth]{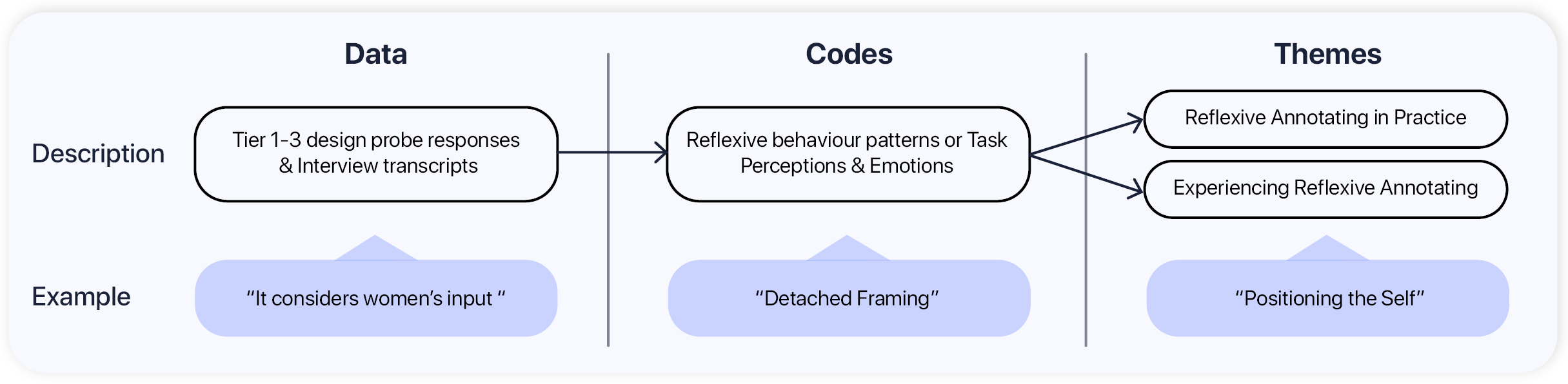}
    \caption{This figure visualises our reflexive thematic analysis as a progression from \emph{data}, to \emph{codes}, to \emph{themes}. The left column represents the materials analysed in the study (participants’ probe responses and interview transcripts). From these materials, we inductively generated codes of recurring patterns in how annotators described their reasoning and feelings towards reflexive annotating. These interpretive codes were then examined and grouped to develop broader themes that represented patterns of shared meaning across participants’ accounts. The example shown in the figure illustrates this movement across levels of abstraction. Codes were developed and revised through repeated engagement with the dataset, and themes emerged through examining relationships across codes. The figure therefore provides a conceptual overview of how we moved from concrete participant accounts to interpretive coding and theme development in our reflexive thematic analysis.}
    \Description{The figure presents a three-stage analytical process arranged from left to right. The left column shows the source data, consisting of participants’ written probe responses and interview transcripts. Arrows indicate a progression to the middle column, which contains inductively generated codes representing recurring patterns in how participants describe their reasoning and experiences with reflexive annotation. A further set of arrows leads to the right column, where these codes are grouped into broader themes that capture shared meanings across participants. An example pathway illustrates how a specific excerpt from the data is interpreted into a code and then contributes to a theme. The layout emphasises an iterative and interpretive process, moving from concrete participant accounts to higher-level thematic understanding.}
    \label{fig:analysis}
\end{figure}

\end{document}